\documentclass[conference]{IEEEtran}
\IEEEoverridecommandlockouts

\usepackage[utf8]{inputenc}
\usepackage[T1]{fontenc}
\usepackage{float}
\usepackage{newtxtext}
\usepackage{newtxmath}

\usepackage{graphicx}
\usepackage{xcolor}
\usepackage{url}
\usepackage{microtype}

\usepackage{array}
\usepackage{booktabs}
\usepackage{colortbl}
\usepackage{makecell}
\usepackage{multirow}
\usepackage{tabularx}
\usepackage{arydshln}
\newcolumntype{Y}{>{\raggedright\arraybackslash}X}
\usepackage{siunitx}

\usepackage{amsmath,amsfonts}

\usepackage{amssymb}
\usepackage{cite}
\usepackage{textcomp}
\usepackage{algorithmic}
\usepackage{etoolbox}

\usepackage{fontawesome5}
\usepackage{tikz} 
\usetikzlibrary{
  shapes.geometric, positioning, arrows.meta, calc, fit,
  backgrounds, shadows.blur
}

\definecolor{Cf}{HTML}{2CA02C}
\definecolor{Cc}{HTML}{1F77B4}
\definecolor{Ci}{HTML}{D62728}
\definecolor{CE}{HTML}{5B9BD5}
\definecolor{CA}{HTML}{70AD47}
\definecolor{CD}{HTML}{C5A55A}
\definecolor{CQ}{HTML}{8C7AB5}
\definecolor{CH}{HTML}{C47D7D}
\definecolor{CT}{HTML}{2C3E50}
\definecolor{CM}{HTML}{7F8C8D}
\definecolor{CG}{HTML}{D4A843}

\tikzset{
  bx/.style n args={3}{
    rectangle, rounded corners=5pt,
    minimum width=#1, minimum height=#2,
    draw=#3!60!black, line width=0.8pt,
    fill=#3!9, text=CT,
    font=\normalfont\footnotesize, align=center, inner sep=4pt,
    blur shadow={shadow blur steps=6, shadow xshift=1pt,
                 shadow yshift=1pt, shadow blur radius=2pt,
                 shadow opacity=0.06}
  },
  bx2/.style n args={3}{
    rectangle, rounded corners=4pt,
    minimum width=#1, minimum height=#2,
    draw=#3!55!black, line width=0.7pt,
    fill=#3!11, text=CT,
    font=\normalfont\fontsize{6.5}{7.5}\selectfont,
    align=center, inner sep=3pt,
    blur shadow={shadow blur steps=5, shadow xshift=0.8pt,
                 shadow yshift=0.8pt, shadow blur radius=1.5pt,
                 shadow opacity=0.05}
  },
  pl/.style={
    rectangle, rounded corners=6pt,
    inner sep=2.5pt, minimum height=11pt,
    font=\normalfont\fontsize{5.5}{6.5}\selectfont\bfseries,
    text=white
  },
  ar/.style={
    -{Stealth[length=4.5pt,width=3.5pt]},
    CT!40, line width=0.8pt,
    shorten >=1.5pt, shorten <=1pt
  },
  ad/.style={
    -{Stealth[length=4.5pt,width=3.5pt]},
    CT!40, line width=0.7pt, densely dashed,
    shorten >=1.5pt, shorten <=1pt
  },
  zn/.style n args={2}{
    rectangle, rounded corners=10pt,
    inner sep=10pt, inner ysep=12pt,
    draw=#1!35!black, line width=0.4pt,
    fill=#1!5, opacity=0.45
  },
}

\newcommand{\tabMapCase}{\textcolor{quantumNavy}{\faHashtag}}          
\newcommand{\tabMapComp}{\textcolor{quantumTeal}{\faFileCode}}        
\newcommand{\tabMapStage}{\textcolor{tensorAmber}{\faStream}}         
\newcommand{\tabMapFig}{\textcolor{quantumPurple}{\faImage}}          
\newcommand{\tabMapDash}{\textcolor{gray}{\faMinus}}                  

\definecolor{stageP1}{HTML}{2563EB}   
\definecolor{stageP2}{HTML}{7C3AED}   
\definecolor{stageP3}{HTML}{059669}   
\definecolor{stageP4}{HTML}{D97706}   
\definecolor{stageP6}{HTML}{DC2626}   
\definecolor{stageP7}{HTML}{0891B2}   

\definecolor{pageMapTintA}{HTML}{F0F4FA}  
\definecolor{pageMapTintB}{HTML}{FEF2F2}  

\definecolor{Tfocus}{HTML}{2CA02C}      
\definecolor{Tcontrib}{HTML}{1F77B4}    
\definecolor{Tinsight}{HTML}{D62728}    
\definecolor{Tedge}{HTML}{5B9BD5}       
\definecolor{Tapp}{HTML}{70AD47}        
\definecolor{Tdata}{HTML}{C5A55A}       
\definecolor{Tqsvc}{HTML}{8C7AB5}       
\definecolor{Tqhw}{HTML}{C47D7D}        
\definecolor{Ttext}{HTML}{2C3E50}       
\definecolor{Tmuted}{HTML}{7F8C8D}      
\definecolor{Taccent}{HTML}{D4A843}     
\definecolor{Tbg}{HTML}{FAFAF7}         

\definecolor{quantumNavy}{HTML}{1A2B4C}
\definecolor{quantumTeal}{HTML}{0E7C7B}
\definecolor{quantumCyan}{HTML}{17BEBB}
\definecolor{quantumPurple}{HTML}{5E2BFF}
\definecolor{tensorAmber}{HTML}{D97706}
\definecolor{archSimTint}{HTML}{F0F4FA} 
\definecolor{archTiTint}{HTML}{F2EEFF}  
\definecolor{archScTint}{HTML}{FEF9E7}  
\definecolor{archNaTint}{HTML}{E8F8F5}  

\definecolor{quantumNavy}{HTML}{1A2B4C}
\definecolor{quantumTeal}{HTML}{0E7C7B}
\definecolor{quantumCyan}{HTML}{17BEBB}
\definecolor{quantumPurple}{HTML}{5E2BFF}
\definecolor{tensorAmber}{HTML}{D97706}
\definecolor{archSimTint}{HTML}{F0F4FA}
\definecolor{archScTint}{HTML}{FEF9E7}
\definecolor{archNaTint}{HTML}{E8F8F5}

\tikzset{
  archbox/.style n args={3}{
    rectangle, rounded corners=6pt,
    minimum width=#1, minimum height=#2,
    draw=#3!65!black, line width=0.9pt,
    fill=#3!10, text=Ttext,
    font=\normalfont\fontsize{7}{8.5}\selectfont,
    align=center, inner sep=5pt,
    blur shadow={shadow blur steps=8, shadow xshift=1.2pt,
                 shadow yshift=1.2pt, shadow blur radius=2.5pt,
                 shadow opacity=0.08}
  },
  tierpill/.style={
    rectangle, rounded corners=10pt,
    inner sep=3.5pt, minimum height=15pt,
    font=\normalfont\fontsize{6}{7}\selectfont\bfseries,
    text=white
  },
  arr/.style={
    -{Stealth[length=5pt,width=3.5pt]},
    Ttext!45, line width=0.9pt,
    shorten >=1.5pt, shorten <=1pt
  },
  arrdash/.style={
    -{Stealth[length=5pt,width=3.5pt]},
    Ttext!45, line width=0.8pt, densely dashed,
    shorten >=1.5pt, shorten <=1pt
  },
  tierzone/.style n args={2}{
    rectangle, rounded corners=14pt,
    inner sep=13pt, inner ysep=15pt,
    draw=#1!40!black, line width=0.5pt,
    fill=#1!6, opacity=0.5
  },
}
\usetikzlibrary{
  shapes.geometric,
  shapes.symbols,
  shapes.misc,
  positioning,
  arrows.meta,
  calc,
  fit,
  backgrounds,
  decorations.pathreplacing,
  shadows.blur,
  patterns,
  matrix
}

\usepackage[utf8]{inputenc}
\usepackage[T1]{fontenc}

\definecolor{UserBg}{HTML}{F5F1EB}
\definecolor{UserBorder}{HTML}{8B7D6B}

\definecolor{NetBg}{HTML}{E8F4F8}
\definecolor{NetBorder}{HTML}{4A90A4}
\definecolor{NetFill}{HTML}{D0E8F0}

\definecolor{AppBg}{HTML}{F0F4E8}
\definecolor{AppBorder}{HTML}{6B8E5A}
\definecolor{AppFill}{HTML}{DCE8D0}

\definecolor{DataBg}{HTML}{F5F0E8}
\definecolor{DataBorder}{HTML}{B89B72}
\definecolor{DataFill}{HTML}{F0E6D0}

\definecolor{QCBg}{HTML}{F0EBF5}
\definecolor{QCBorder}{HTML}{7A6FA8}
\definecolor{QCFill}{HTML}{E0D8F0}

\definecolor{HWBg}{HTML}{F5EBEB}
\definecolor{HWBorder}{HTML}{A67070}
\definecolor{HWFill}{HTML}{F0D8D8}

\definecolor{TextDark}{HTML}{2C3E50}
\definecolor{TextLight}{HTML}{FFFFFF}
\definecolor{TextMuted}{HTML}{6B7B8D}
\definecolor{AccentLine}{HTML}{C4A265}
\definecolor{ArrowColor}{HTML}{546E7A}

\tikzset{
  svc/.style n args={3}{
    rectangle, rounded corners=4pt,
    minimum width=#1, minimum height=#2,
    draw=#3, line width=0.8pt,
    font=\fontsize{7.5}{9}\selectfont,
    align=center, text=TextDark,
    inner sep=4pt,
    blur shadow={shadow blur steps=6, shadow xshift=1pt, shadow yshift=1pt,
                 shadow blur radius=2pt, shadow opacity=0.08}
  },
  tierlabel/.style={
    rectangle, rounded corners=8pt,
    inner sep=3pt,
    font=\fontsize{6.5}{8}\selectfont\bfseries,
    text=TextLight,
    minimum height=14pt
  },
  flow/.style={
    -{Stealth[length=5pt,width=4pt]},
    ArrowColor, line width=0.9pt,
    shorten >=1.5pt, shorten <=1pt
  },
  flowdashed/.style={
    -{Stealth[length=5pt,width=4pt]},
    ArrowColor, line width=0.7pt, densely dashed,
    shorten >=1.5pt, shorten <=1pt
  },
  tierbg/.style n args={2}{
    rectangle, rounded corners=10pt,
    inner sep=10pt, inner ysep=14pt,
    draw=#1, line width=0.6pt,
    fill=#2,
    opacity=0.35
  },
}

\definecolor{quantumNavy}{HTML}{1A2B4C}      
\definecolor{quantumTeal}{HTML}{0E7C7B}      
\definecolor{quantumCyan}{HTML}{17BEBB}      
\definecolor{quantumPurple}{HTML}{5E2BFF}    

\definecolor{cpuBlue}{HTML}{2563EB}          
\definecolor{gpuGreen}{HTML}{059669}         
\definecolor{tensorAmber}{HTML}{D97706}      

\definecolor{awsRowTint}{HTML}{F0F4FA}       
\definecolor{cudaRowTint}{HTML}{E8F8F5}      
\definecolor{headerWhite}{HTML}{FFFFFF}      

\newcolumntype{L}[1]{>{\raggedright\arraybackslash}p{#1}}
\newcolumntype{C}[1]{>{\centering\arraybackslash}p{#1}}
\newcolumntype{R}[1]{>{\raggedleft\arraybackslash}p{#1}}

\newcommand{\cpuicon}{\textcolor{cpuBlue}{\faMicrochip}}
\newcommand{\gpuicon}{\textcolor{gpuGreen}{\faServer}}
\newcommand{\awsicon}{\textcolor{quantumTeal}{\faCloud}}
\newcommand{\quantumicon}{\textcolor{quantumPurple}{\faAtom}}
\newcommand{\nvidiaicon}{\textcolor{gpuGreen}{\faVideo}}
\newcommand{\chipicon}{\textcolor{quantumCyan}{\faMemory}}
\newcommand{\bolticon}{\textcolor{tensorAmber}{\faBolt}}

\newcommand{\devicename}[1]{\textbf{\textcolor{quantumNavy}{#1}}}

\newcommand{\cpubadge}{%
  \tikz[baseline=-2pt]{%
    \node[rounded corners=3pt,fill=cpuBlue!10,draw=cpuBlue!60,line width=0.5pt,inner sep=3pt,font=\normalfont\small\bfseries] {\cpuicon~CPU};%
  }%
}
\newcommand{\gpubadge}[1]{%
  \tikz[baseline=-2pt]{%
    \node[rounded corners=3pt,fill=gpuGreen!10,draw=gpuGreen!60,line width=0.5pt,inner sep=3pt,font=\normalfont\small\bfseries] {\gpuicon~GPU #1};%
  }%
}

\newcommand{\qubitcount}[1]{\textbf{\textcolor{quantumNavy}{#1}}}

\newcommand{\cicon}[2]{\textcolor{#1}{#2}} 
\newcommand{\sdg}[2]{\textcolor{#1}{\textbf{#2}}} 
\newcommand{\comm}[2]{\textcolor{#1}{#2}} 

\definecolor{tagFocus}{HTML}{2CA02C}    
\definecolor{tagContrib}{HTML}{1F77B4}  
\definecolor{tagInsight}{HTML}{D62728}  

\newcommand{\tagF}{\textcolor{tagFocus}{\textbf{F}}}
\newcommand{\tagC}{\textcolor{tagContrib}{\textbf{C}}}
\newcommand{\tagI}{\textcolor{tagInsight}{\textbf{I}}}
\definecolor{cResearch}{HTML}{1F77B4}   
\definecolor{cEng}{HTML}{FF7F0E}        
\definecolor{cDesign}{HTML}{9467BD}     
\definecolor{cEdu}{HTML}{2CA02C}        
\definecolor{cInvest}{HTML}{8C564B}     
\definecolor{cPolicy}{HTML}{D62728}     
\definecolor{cPublic}{HTML}{7F7F7F}     
\definecolor{cSustain}{HTML}{17BECF}    

\definecolor{sdg4}{HTML}{C5192D}
\definecolor{sdg9}{HTML}{FD6925}
\definecolor{sdg12}{HTML}{BF8B2E}
\definecolor{sdg13}{HTML}{3F7E44}
\definecolor{sdg16}{HTML}{00689D}
\definecolor{sdg17}{HTML}{19486A}
\newcommand{\vicons}[1]{\begin{tabular}[c]{@{}c@{}}#1\end{tabular}}

\usetikzlibrary{positioning,arrows.meta,fit,calc}

\definecolor{cEdge}{HTML}{2563EB}     
\definecolor{cApi}{HTML}{0EA5E9}      
\definecolor{cApp}{HTML}{10B981}      
\definecolor{cData}{HTML}{F59E0B}     
\definecolor{cQ}{HTML}{8B5CF6}        
\definecolor{cObj}{HTML}{64748B}      
\definecolor{cLine}{HTML}{334155}     
\definecolor{cBg}{HTML}{F8FAFC}       

\tikzset{
  box/.style={
    draw=cLine, rounded corners=2pt, line width=0.4pt,
    fill=cBg, inner sep=4pt, align=left, font=\footnotesize
  },
  arrow/.style={-{Latex[length=2mm,width=1.5mm]}, line width=0.55pt, draw=cLine},
  tag/.style={font=\scriptsize, text=cLine},
  ic/.style={font=\footnotesize}
}

\def\BibTeX{{\rm B\kern-.05em{\sc i\kern-.025em b}\kern-.08em
    T\kern-.1667em\lower.7ex\hbox{E}\kern-.125emX}}

\usepackage{tcolorbox}
\tcbuselibrary{skins,breakable}

\usepackage[
  colorlinks=true,
  linkcolor=quantumNavy,
  citecolor=darkgray,
  urlcolor=quantumTeal,
  bookmarks=true,
  unicode=true
]{hyperref}

\begin{document}
\title{Quantum Futures Interactive: A Live Demonstration of Post-Quantum Blockchain Security, Infrastructure Tradeoffs, and Sustainable Distributed Trust}

\author{

  \IEEEauthorblockN{Dongping Liu\textsuperscript{\dag}\textsuperscript{\ddag}}
  \IEEEauthorblockA{\textit{Amazon Web Services}\\
    Hong Kong, China\\}
  \and
  \IEEEauthorblockN{Aoyu Zhang\textsuperscript{\dag}}
  \IEEEauthorblockA{\textit{Amazon Web Services}\\
    Beijing, China\\}
  \and
  \IEEEauthorblockN{Luyao Zhang\textsuperscript{\dag}\textsuperscript{*}}
  \IEEEauthorblockA{\textit{Duke Kunshan University}\\
    Suzhou, China\\}
  \thanks{\textsuperscript{*}Corresponding author: Luyao Zhang (lz183@duke.edu), Digital Innovation Research Center and Social Science Division, Duke Kunshan University. Address: Duke Avenue No.8, Kunshan, Suzhou, Jiangsu, China, 215316. \textsuperscript{\dag} Equal contributions. Authors are listed in alphabetical order by last name and then first name.  \textsuperscript{\ddag} Work done while at Amazon Web Services. Dongping Liu is currently with Tenorshare, Hong Kong, China.}
}

\maketitle

\begin{abstract}
Advances in quantum computing introduce long-term security challenges for widely deployed public-key cryptographic systems used across blockchain platforms and decentralized applications. Although post-quantum cryptography (PQC) standards are emerging, understanding quantum risk remains fragmented across research, engineering, governance, and investment communities. This demo presents \textit{Quantum Futures Interactive}, a live interdisciplinary demonstration platform combining educational visualization, participatory interaction, and cryptographic artifact generation to illustrate the transition from classical to quantum-resilient blockchain systems. Participants engage in a structured interaction flow including quantum threat education, sentiment capture, technology prioritization, infrastructure tradeoff exploration, and generation of demonstrative post-quantum artifacts using a toy LWE-based construction. The system integrates distributed trust concepts, sustainability-aware infrastructure considerations, and responsible innovation within an interactive decision framework. The demonstration supports interdisciplinary dialogue on blockchain resilience while aligning with United Nations Sustainable Development Goals (SDGs).
\end{abstract}

\begin{IEEEkeywords}
Post-Quantum Cryptography, Blockchain Security, Interactive Demonstration, Sustainable
Computing, Distributed Trust, Quantum Randomness
\end{IEEEkeywords}

\begin{table}[!t]
\centering
\caption{Demonstration Flow, Technical Contributions, and Community Impact with UN SDG Alignment}
\label{tab:interdisciplinary_onecol}
\scriptsize

\setlength{\aboverulesep}{0.5pt}
\setlength{\belowrulesep}{0.5pt}
\setlength{\tabcolsep}{0.1pt}
\renewcommand{\arraystretch}{0.88}
\setlength{\dashlinedash}{0.25pt}
\setlength{\dashlinegap}{0.8pt}

\begin{tabular}{@{}>{\centering\arraybackslash}m{0.48cm}
                >{\centering\arraybackslash}m{0.92cm}
                >{\raggedright\arraybackslash}m{4.65cm}
                >{\centering\arraybackslash}m{1.58cm}
                >{\centering\arraybackslash}m{0.55cm}@{}}
\toprule
\textbf{Case} &
\textbf{Stage} &
\textbf{\tagF: Focus \ \tagC: Contribution \ \tagI: Insight} &
\textbf{Communities} &
\textbf{SDGs} \\
\midrule

0 &
\cicon{cEdu}{\faRocket}\textbf{P1} &
\tagF: Introduce PQC relevance to blockchain cryptography.
\tagC: Establish shared language linking quantum advances and blockchain trust assumptions.
\tagI: Prepare participants for quantum-safe technical concepts. &
\vicons{\comm{cResearch}{\faFlask}\comm{cDesign}{\faPalette}\\
\comm{cEdu}{\faChalkboardTeacher}\comm{cInvest}{\faChartLine}} &
\multirow{2}{*}{\vicons{\sdg{sdg4}{4}\\\sdg{sdg9}{9}\\\sdg{sdg16}{16}}} \\

1 &
\cicon{cResearch}{\faAtom}\textbf{P2} &
\tagF: Explain quantum threats and migration motivation.
\tagC: Connect algorithmic advances to risks for blockchain signatures and data integrity.
\tagI: Show how discovery informs engineering and governance decisions. &
\vicons{\comm{cResearch}{\faFlask}\comm{cEng}{\faCogs}\\
\comm{cPolicy}{\faBalanceScale}} &
\\

\hdashline

2 &
\cicon{cPolicy}{\faCheckCircle}\textbf{P3} &
\tagF: Transition from learning to informed participation.
\tagC: Apply consent and responsible handling within a cryptographic workflow.
\tagI: Reinforce transparency as a basis for trusted infrastructure. &
\vicons{\comm{cDesign}{\faPalette}\comm{cPolicy}{\faGavel}\\
\comm{cEdu}{\faChalkboardTeacher}} &
\multirow{3}{*}{\vicons{\sdg{sdg4}{4}\\\sdg{sdg9}{9}\\\sdg{sdg16}{16}\\\sdg{sdg17}{17}}} \\

3 &
\cicon{cPublic}{\faCommentDots}\textbf{P4} &
\tagF: Capture public perception with minimal effort.
\tagC: Contrast intuition with technical constraints of quantum-safe migration.
\tagI: Reveal communication gaps affecting adoption. &
\vicons{\comm{cPublic}{\faUsers}\comm{cEdu}{\faChalkboardTeacher}\\
\comm{cPolicy}{\faBalanceScale}\comm{cResearch}{\faFlask}} &
\\

&
\cicon{cInvest}{\faChartBar}\textbf{P5} &
\tagF: Display aggregated sentiment and preferences.
\tagC: Elicit priorities across quantum-safe infrastructure options.
\tagI: Illustrate ecosystem consensus formation. &
\vicons{\comm{cInvest}{\faChartLine}\comm{cPolicy}{\faBalanceScale}\\
\comm{cEng}{\faCogs}\comm{cResearch}{\faFlask}} &
\\

\hdashline

4--5 &
\cicon{cSustain}{\faVial}\textbf{P6} &
\tagF: Explore quantum execution environments via simulation.
\tagC: Compare architectures in performance, availability, and sustainability.
\tagI: Link infrastructure choices to strategic tradeoffs. &
\vicons{\comm{cEng}{\faCogs}\comm{cInvest}{\faChartLine}\\
\comm{cSustain}{\faLeaf}\comm{cResearch}{\faFlask}} &
\multirow{2}{*}{\vicons{\sdg{sdg4}{4}\\\sdg{sdg9}{9}\\\sdg{sdg12}{12}\\\sdg{sdg13}{13}\\\sdg{sdg16}{16}}} \\

6--7 &
\cicon{cEng}{\faLock}\textbf{P7} &
\tagF: Generate a demonstrative post-quantum artifact using a toy LWE-based construction.
\tagC: Provide execution metadata linking device selection and provenance.
\tagI: Connect learning to verifiable trust outputs. &
\vicons{\comm{cEng}{\faCogs}\comm{cResearch}{\faFlask}\\
\comm{cEdu}{\faChalkboardTeacher}\comm{cInvest}{\faChartLine}} &
\\

\bottomrule
\end{tabular}
\vspace{0.58mm}
\scriptsize
\noindent\parbox{\columnwidth}{\raggedright
\textit{Table notes:}

The live system implements the workflow across eight React component pages
(case indices 0--7 in \texttt{page.tsx}):
\texttt{NobelPrizePage},
\texttt{QuantumBlockchainPage},
\texttt{QuantumReadyPage},
\texttt{SentimentPage},
\texttt{SimulatorSelectionPage},
\texttt{QPUSelectionPage},
\texttt{KeyGenerationPage}, and
\texttt{CompletionPage}.
The seven-stage conceptual grouping merges cases 4--5 (simulator and QPU
selection) into P6 ``Infrastructure Tradeoffs'' and absorbs case 7
(completion) into P7 ``Artifact Generation.''

\tagF\ indicates focus.
\tagC\ indicates contribution.
\tagI\ indicates intended insight.
\comm{cResearch}{\faFlask} denotes researchers and scientific contributors.
\comm{cEng}{\faCogs} denotes engineers and technical practitioners.
\comm{cDesign}{\faPalette} denotes designers and user-experience specialists.
\comm{cEdu}{\faChalkboardTeacher} denotes educators and learning facilitators.
\comm{cInvest}{\faChartLine} denotes investors and strategic decision makers.
\comm{cPolicy}{\faBalanceScale} denotes governance and regulatory perspectives.
\comm{cPolicy}{\faGavel} denotes policy and institutional authority.
\comm{cPublic}{\faUsers} denotes public participants and non-specialist audiences.
\comm{cSustain}{\faLeaf} denotes sustainability and environmental considerations.
SDG~4: Quality Education.
SDG~9: Industry, Innovation and Infrastructure.
SDG~12: Responsible Consumption and Production.
SDG~13: Climate Action.
SDG~16: Peace, Justice and Strong Institutions.
SDG~17: Partnerships for the Goals.
}
\vspace{-1.8mm}

\end{table}

\section{Demonstration Motivation and System Overview}

Public-key cryptography enables blockchain systems to establish identity, validate transactions, and maintain distributed trust without centralized control \cite{Yaga2018BlockchainOverview}, enabling blockchain systems to function as trust infrastructure for global digital exchange beyond distributed databases\cite{Budish2025TrustAtScale}. As economic coordination expands into intelligent economies, metaverse environments, and digital twin systems, the reliability of blockchain-based infrastructures \cite{9815155} depends directly on the security assumptions of their underlying cryptographic primitives. Advances in quantum computing challenge the hardness assumptions supporting widely deployed public-key schemes, motivating transition toward post-quantum cryptographic (PQC) mechanisms \cite{Bernstein2017PostQuantum,nistpqc}. Despite substantial progress in post-quantum algorithm design and standardization \cite{Joseph2022PostQuantum,10048976}, migration toward quantum-safe blockchain infrastructure introduces challenges beyond cryptographic implementation \cite{cryptoeprint:2025/1626}. Deployment decisions increasingly involve infrastructure readiness, sustainability considerations, governance constraints, and ecosystem coordination among developers, operators, regulators, and users \cite{11059920}. Cryptographic mechanisms, infrastructure tradeoffs, and adoption dynamics are rarely demonstrated together in interactive environments accessible to both technical and non-technical stakeholders, despite their importance for understanding technological transitions across interdisciplinary communities \cite{10.1145/3555591,10310184}.

We present \textit{Quantum Futures Interactive}, a live demonstration system that allows participants to experience quantum risk awareness education, post-quantum blockchain security examination, and infrastructure decision-making through interactive visualization and simulation. The demonstration contributes three implementation-oriented components.
First, the system architecture, illustrated in Fig.~\ref{fig:architecture}, presents an integrated serverless design connecting user interaction, application services, persistent storage, and quantum execution within a unified framework. Second, the implementation is released as an open-source platform supporting reproducibility and continued research development, available on GitHub at \url{https://github.com/QuantBlockchain/qc-bc-interactive}. Third, the demonstration evaluates the impact of cryptographic transition across stakeholder communities, as summarized in Table~\ref{tab:interdisciplinary_onecol}, linking infrastructure decisions to interdisciplinary engagement and societal objectives aligned with the United Nations Sustainable Development Goals\cite{Carlsen2022SDGStatus}. 

This demonstration extends the pedagogical quantum-cryptography framework introduced
in QSignAI~\cite{qsignai2025}, re-purposing its ToyLWE-Quantum-Seeded-Demo module
toward blockchain-security education. Where QSignAI targets AI-bot and social-platform
deployment scenarios, the present work focuses on infrastructure-tradeoff education
across quantum simulators and QPUs within an interactive blockchain-security context. Complementary work in this research program explores quantum-computing
hardware through interactive generative world models
\cite{zhang2026quantumcinemainteractivecinematic} and develops
cost-aware evaluation methods for visual AI agents performing quantum
code generation
\cite{liu2026quantumcircuitvisioncostaware}.

\section{System Architecture and Implementation}
\label{sec:architecture}
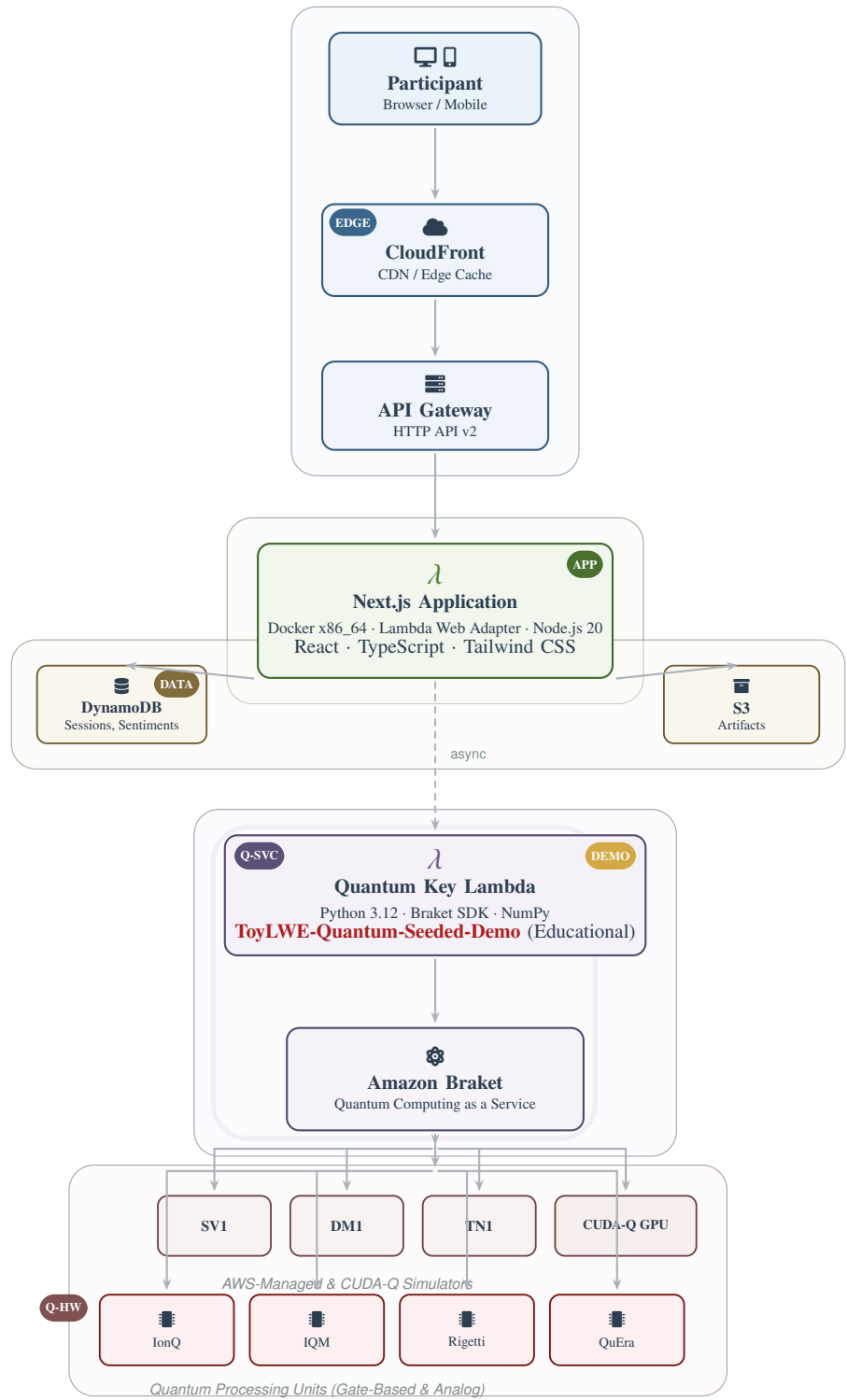
\begin{figure*}[!t]
\centering
\begin{tikzpicture}[
  node distance=1.4cm and 1.8cm,
  every node/.style={font=\sffamily}
]

\pgfdeclarelayer{bg}
\pgfdeclarelayer{mid}
\pgfsetlayers{bg,mid,main}

\node[bx={3.0cm}{1.3cm}{CE}, fill=CE!13] (U)
  {\faIcon{desktop}~\faIcon{mobile-alt}\\[1pt]
   \textbf{Participant}\\[-1pt]
   \fontsize{6}{7}\selectfont Browser / Mobile};

\node[bx={3.2cm}{1.3cm}{CE}, fill=CE!10, below=1.1cm of U] (CF)
  {\faIcon{cloud}\\[1pt]
   \textbf{CloudFront}\\[-1pt]
   \fontsize{6}{7}\selectfont CDN / Edge Cache};

\node[pl, fill=CE!65!black, anchor=north west, xshift=3pt, yshift=-2pt]
  at (CF.north west) {\fontsize{5}{6}\selectfont EDGE};

\node[bx={3.2cm}{1.25cm}{CE}, fill=CE!10, below=0.9cm of CF] (AG)
  {\faIcon{server}\\[1pt]
   \textbf{API Gateway}\\[-1pt]
   \fontsize{6}{7}\selectfont HTTP API v2};

\node[bx={5.0cm}{1.9cm}{CA}, fill=CA!10, below=1.3cm of AG] (APP)
  {\textcolor{CA!75!black}{\large$\lambda$}\\[1pt]
   \textbf{Next.js Application}\\[1pt]
   \fontsize{6.5}{7.5}\selectfont
   Docker x86\_64 · Lambda Web Adapter · Node.js 20\\[-1pt]
   React · TypeScript · Tailwind CSS};

\node[pl, fill=CA!65!black, anchor=north east, xshift=-4pt, yshift=-3pt]
  at (APP.north east) {\fontsize{5}{6}\selectfont APP};

\node[bx2={2.4cm}{1.1cm}{CD}, fill=CD!10,
      below left=-0.2cm and 0.7cm of APP] (DB)
  {\faIcon{database}\\[1pt]
   \textbf{DynamoDB}\\[-1pt]
   \fontsize{5.5}{6.5}\selectfont Sessions, Sentiments};

\node[pl, fill=CD!65!black, anchor=north east, xshift=-3pt, yshift=-2pt]
  at (DB.north east) {\fontsize{5}{6}\selectfont DATA};

\node[bx2={2.2cm}{1.1cm}{CD}, fill=CD!10,
      below right=-0.2cm and 0.7cm of APP] (S3n)
  {\faIcon{archive}\\[1pt]
   \textbf{S3}\\[-1pt]
   \fontsize{5.5}{6.5}\selectfont Artifacts};

\node[bx={4.8cm}{1.7cm}{CQ}, fill=CQ!10, below=2.2cm of APP] (QK)
  {\textcolor{CQ!75!black}{\large$\lambda$}\\[1pt]
   \textbf{Quantum Key Lambda}\\[1pt]
   \fontsize{6.5}{7.5}\selectfont
   Python 3.12 · Braket SDK · NumPy\\[-1pt]
   \textcolor{Ci!80!black}{\textbf{ToyLWE-Quantum-Seeded-Demo}}~(Educational)};

\node[pl, fill=CG, text=white, anchor=north east, xshift=-4pt, yshift=-3pt]
  at (QK.north east) {\fontsize{5}{6}\selectfont DEMO};
\node[pl, fill=CQ!65!black, anchor=north west, xshift=4pt, yshift=-3pt]
  at (QK.north west) {\fontsize{5}{6}\selectfont Q-SVC};

\node[bx={4.2cm}{1.45cm}{CQ}, fill=CQ!8, below=1.0cm of QK] (BK)
  {\faIcon{atom}\\[1pt]
   \textbf{Amazon Braket}\\[-1pt]
   \fontsize{6}{7}\selectfont Quantum Computing as a Service};

\node[bx2={1.6cm}{0.85cm}{CH}, fill=CH!10,
      below left=0.9cm and 0.2cm of BK] (SV1)
  {\fontsize{6}{7}\selectfont\textbf{SV1}};
\node[bx2={1.6cm}{0.85cm}{CH}, fill=CH!10, right=0.25cm of SV1] (DM1)
  {\fontsize{6}{7}\selectfont\textbf{DM1}};
\node[bx2={1.6cm}{0.85cm}{CH}, fill=CH!10, right=0.25cm of DM1] (TN1)
  {\fontsize{6}{7}\selectfont\textbf{TN1}};
\node[bx2={2.0cm}{0.85cm}{CH}, fill=CH!13, right=0.25cm of TN1] (CQG)
  {\fontsize{5.5}{6.5}\selectfont\textbf{CUDA-Q GPU}};

\node[bx2={1.9cm}{1.0cm}{Ci}, fill=Ci!7,
      below=2.3cm of BK, xshift=-3.8cm] (IONQ)
  {\faIcon{microchip}\\[1pt]
   \fontsize{5.5}{6.5}\selectfont IonQ};
\node[bx2={1.9cm}{1.0cm}{Ci}, fill=Ci!7, right=0.2cm of IONQ] (IQM)
  {\faIcon{microchip}\\[1pt]
   \fontsize{5.5}{6.5}\selectfont IQM};
\node[bx2={1.9cm}{1.0cm}{Ci}, fill=Ci!7, right=0.2cm of IQM] (RIG)
  {\faIcon{microchip}\\[1pt]
   \fontsize{5.5}{6.5}\selectfont Rigetti};
\node[bx2={1.9cm}{1.0cm}{Ci}, fill=Ci!7, right=0.2cm of RIG] (QUE)
  {\faIcon{microchip}\\[1pt]
   \fontsize{5.5}{6.5}\selectfont QuEra};

\node[pl, fill=CH!65!black, anchor=north east, xshift=-0.15cm, yshift=0pt]
  at (IONQ.north west) {\fontsize{5}{6}\selectfont Q-HW};

\node[font=\sffamily\fontsize{6}{7}\selectfont\itshape, text=CM,
      below=0.18cm of DM1.south]
  {AWS-Managed \& CUDA-Q Simulators};

\node[font=\sffamily\fontsize{6}{7}\selectfont\itshape, text=CM,
      below=0.12cm of IQM.south]
  {Quantum Processing Units (Gate-Based \& Analog)};

\begin{pgfonlayer}{mid}
  \node[zn={CE}{CE!3}, fit=(U)(CF)(AG),
        inner xsep=12pt, inner ysep=10pt] {};
  \node[zn={CA}{CA!3}, fit=(APP),
        inner xsep=12pt, inner ysep=10pt] {};
  \node[zn={CD}{CD!3}, fit=(DB)(S3n),
        inner xsep=10pt, inner ysep=10pt] {};
  \node[zn={CQ}{CQ!3}, fit=(QK)(BK),
        inner xsep=12pt, inner ysep=10pt] {};
  \node[zn={CH}{CH!3},
        fit=(SV1)(DM1)(TN1)(CQG)(IONQ)(IQM)(RIG)(QUE),
        inner sep=12pt] {};
\end{pgfonlayer}

\begin{pgfonlayer}{bg}
  \draw[CQ!15, line width=2pt, rounded corners=14pt]
    ($(QK.north west)+(-0.15,0.1)$) rectangle
    ($(BK.south east)+(0.15,-0.1)$);
\end{pgfonlayer}

\draw[ar] (U) -- (CF);
\draw[ar] (CF) -- (AG);
\draw[ar] (AG) -- (APP);
\draw[ar] (APP.south west) -- (DB.north);
\draw[ar] (APP.south east) -- (S3n.north);
\draw[ad] (APP.south) -- (QK.north)
  node[midway, right=2.5pt,
       font=\sffamily\fontsize{5.5}{6.5}\selectfont, text=CM]
    {async};
\draw[ar] (QK) -- (BK);

\coordinate (dropSim) at ($(BK.south)+(0,-7pt)$);
\draw[ar] (BK.south) -- (dropSim);
\draw[ar] (dropSim) -| (SV1.north);
\draw[ar] (dropSim) -| (DM1.north);
\draw[ar] (dropSim) -| (TN1.north);
\draw[ar] (dropSim) -| (CQG.north);

\coordinate (dropQpu) at ($(BK.south)+(0,-16pt)$);
\draw[ar] (BK.south) -- (dropQpu);
\draw[ar] (dropQpu) -| (IONQ.north);
\draw[ar] (dropQpu) -| (IQM.north);
\draw[ar] (dropQpu) -| (RIG.north);
\draw[ar] (dropQpu) -| (QUE.north);

\end{tikzpicture}
\caption{System architecture of Quantum Futures Interactive.
  Participants connect through CloudFront to a serverless pipeline:
  API Gateway routes requests to a Dockerized Next.js application
  running on AWS Lambda (x86\_64). Application state persists in
  DynamoDB; optional artifacts are stored in S3. Quantum execution
  is delegated via a dedicated Python Lambda invoking Amazon Braket
  simulators (SV1, DM1, TN1, and NVIDIA CUDA-Q GPU) or physical QPUs
  (IonQ, IQM, Rigetti, QuEra). The signature module
  (ToyLWE-Quantum-Seeded-Demo) carries an explicit educational-use
  warning and is not intended for production deployment.}
\label{fig:architecture}
\end{figure*}
Figure~\ref{fig:architecture} illustrates the system architecture, implemented as a serverless pipeline integrating user interaction, application services, data persistence, and quantum execution. The architecture follows serverless computing principles where event-driven functions support scalability, reproducibility, and low operational overhead for live demonstrations~\cite{9756233}. Client requests flow through CloudFront and API Gateway to a Lambda-hosted Next.js application (Docker x86\_64) that manages the interactive experience and API logic. Application state persists in DynamoDB; optional artifacts are stored in S3.

Quantum execution is delegated asynchronously to a dedicated Python Lambda invoking Amazon Braket simulators or quantum processing units (QPUs), reflecting emerging quantum-computing-as-a-service architectures~\cite{ahmad2023engineeringsoftwaresystemsquantum,Gonzalez2021AmazonBraket}. Table~\ref{tab:simulators} lists the simulator options available to participants---AWS-managed simulators (SV1, DM1, TN1) and the NVIDIA CUDA-Q GPU simulator---summarizing their simulation types, qubit capacities, and acceleration backends. Table~\ref{tab:qpus} presents the QPU alternatives, spanning trapped-ion, superconducting, and neutral-atom analog architectures, together with representative devices and evaluation metrics. The full device specifications appear in Appendix~\ref{app:simulators}--\ref{app:qpus}. 

Participants are informed that simulator-based execution does not produce
physical quantum randomness---outcomes are computed rather than measured---while QPU
backends generate probabilistic measurement statistics from physical quantum
superposition. Table~\ref{tab:tradeoffs} summarizes the illustrative tradeoff dimensions
across all backends.
\begin{table}[t]
\centering
\caption{Illustrative infrastructure tradeoff dimensions across quantum execution
backends. Task times are representative order-of-magnitude ranges; energy profiles
are drawn from quantum computing life-cycle assessment~\cite{Cordier2025EnvironmentalLCAQuantum}.}
\label{tab:tradeoffs}
\footnotesize
\setlength{\tabcolsep}{3pt}
\renewcommand{\arraystretch}{0.9}
\begin{tabular}{@{}lcccc@{}}
\toprule
\textbf{Backend} & \textbf{Type} & \textbf{Task time} & \textbf{Physical} & \textbf{Energy note} \\
 & & \textbf{(repr.)} & \textbf{randomness?} & \\
\midrule
SV1 (sim.) & State-vector & Seconds & No (pseudo) & Conventional DC \\
DM1 (sim.) & Density matrix & Seconds & No (pseudo) & Conventional DC \\
TN1 (sim.) & Tensor network & Seconds & No (pseudo) & GPU-accelerated \\
CUDA-Q GPU & CUDA-Q sim. & Seconds & No (pseudo) & NVIDIA GPU \\
\hdashline
IonQ & Trapped-ion & Sec.--min. & Yes & Laser + UHV \\
IQM & Superconducting & Sec.--min. & Yes & mK cryogenics \\
Rigetti & Superconducting & Sec.--min. & Yes & mK cryogenics \\
QuEra & Neutral-atom & Sec.--min. & Yes & Optical trapping \\
\bottomrule
\end{tabular}
\end{table}
The signature module (\texttt{ToyLWE-Quantum-Seeded-Demo}) uses deliberately reduced LWE parameters for educational demonstration only; it is explicitly labeled as not for production use and does not provide security guarantees equivalent to standards-grade schemes such as ML-DSA or Falcon.

\begin{table*}[!htbp]
\centering
\caption{Quantum computing simulators presented in Page 6. The table lists
AWS-managed simulators and the NVIDIA CUDA-Q GPU simulator; additional
simulator options are available through the Amazon Braket SDK.}
\label{tab:simulators}

\renewcommand{\arraystretch}{1.5}
\setlength{\tabcolsep}{12pt}

\begin{tabular}{@{}llcl@{}}
\toprule
\rowcolor{quantumNavy}
\textcolor{white}{\faAtom~\textbf{Device}} &
\textcolor{white}{\textbf{Simulation Type}} &
\textcolor{white}{\textbf{Max Qubits}} &
\textcolor{white}{\textbf{Acceleration}} \\
\midrule
\rowcolor{awsRowTint}
\awsicon~\devicename{AWS SV1} &
\chipicon~State Vector &
\qubitcount{34} &
\cpubadge \\
\rowcolor{white}
\awsicon~\devicename{AWS DM1} &
\quantumicon~Density Matrix &
\qubitcount{17} &
\cpubadge \\
\rowcolor{awsRowTint}
\awsicon~\devicename{AWS TN1} &
\bolticon~Tensor Network &
\qubitcount{50} &
\gpubadge{(cuQuantum)} \\
\midrule
\rowcolor{cudaRowTint}
\nvidiaicon~\devicename{CUDA-Q GPU} &
\chipicon~CUDA-Q State Vector &
\qubitcount{30+} &
\gpubadge{(NVIDIA)} \\
\bottomrule
\end{tabular}
\label{app:simulators}
\end{table*}


\definecolor{quantumNavy}{HTML}{1A2B4C}
\definecolor{quantumTeal}{HTML}{0E7C7B}
\definecolor{quantumCyan}{HTML}{17BEBB}
\definecolor{quantumPurple}{HTML}{5E2BFF}
\definecolor{tensorAmber}{HTML}{D97706}

\colorlet{archSimTint}{quantumTeal!8}
\colorlet{archTiTint}{quantumPurple!8}
\colorlet{archScTint}{tensorAmber!10}
\colorlet{archNaTint}{quantumCyan!8}

\newcommand{\tabQpusClassical}{\textcolor{quantumTeal}{\faCloud}}           
\newcommand{\tabQpusTrappedIon}{\textcolor{quantumPurple}{\faAtom}}         
\newcommand{\tabQpusSuperconducting}{\textcolor{tensorAmber}{\faMicrochip}} 
\newcommand{\tabQpusNeutralAtom}{\textcolor{quantumCyan}{\faEye}}           

\newcommand{\tabQpusRowA}{\textcolor{quantumNavy}{\textbf{Classical Simulator}}}
\newcommand{\tabQpusRowB}{\textcolor{quantumNavy}{\textbf{Trapped-Ion QPU}}}
\newcommand{\tabQpusRowC}{\textcolor{quantumNavy}{\textbf{Superconducting QPU}}}
\newcommand{\tabQpusRowD}{\textcolor{quantumNavy}{\textbf{Neutral-Atom Analog}}}

\begin{table*}[!htbp]
\centering
\caption{Quantum computing architectures presented in Page 7, including
representative devices and evaluation metrics considered during device
selection.}
\label{tab:qpus}

\renewcommand{\arraystretch}{1.55}
\setlength{\tabcolsep}{10pt}

\begin{tabular}{@{}>{\raggedright}p{3.0cm}>{\raggedright}p{4.2cm}>{\raggedright}p{2.7cm}>{\raggedright\arraybackslash}p{3.8cm}@{}}
\toprule
\rowcolor{quantumNavy}
\textcolor{white}{\faAtom~\textbf{Architecture}} &
\textcolor{white}{\textbf{Definition}} &
\textcolor{white}{\textbf{Example Devices}} &
\textcolor{white}{\textbf{Evaluation Metrics}} \\
\midrule
\rowcolor{archSimTint}
\tabQpusClassical~\tabQpusRowA &
Classical emulation of quantum circuits &
AWS SV1, AWS DM1, AWS TN1 &
Capacity, reproducibility, runtime \\
\addlinespace[2pt]
\rowcolor{archTiTint}
\tabQpusTrappedIon~\tabQpusRowB &
Physical qubits in trapped atomic ions &
IonQ Aria, IonQ Forte &
Fidelity, connectivity \\
\addlinespace[2pt]
\rowcolor{archScTint}
\tabQpusSuperconducting~\tabQpusRowC &
Artificial atoms via Josephson junctions &
IQM Garnet, Rigetti Ankaa-3 &
Qubit count, gate speed \\
\addlinespace[2pt]
\rowcolor{archNaTint}
\tabQpusNeutralAtom~\tabQpusRowD &
Optically trapped neutral atoms &
QuEra Aquila &
Measurement fidelity \\
\bottomrule
\end{tabular}
\label{app:qpus}
\end{table*}

\section{Demonstration Flow and Community Impact}
\label{sec:demo-flow}

The demonstration follows a seven-stage interaction flow summarized in Table~\ref{tab:interdisciplinary_onecol}\footnote{The live implementation comprises eight React component pages (cases 0--7 in \texttt{page.tsx}); the seven-stage narrative groups simulator and QPU selection into a single ``Infrastructure Tradeoffs'' stage (P6) and absorbs the \texttt{CompletionPage} into the artifact-generation stage (P7).}. Table~\ref{tab:page-mapping} provides the complete mapping between implemented React pages, conceptual stages, and appendix figure references.


\begin{table*}[!htbp]
\small
\centering
\caption{Mapping between implementation pages, conceptual stages, and
appendix figures.}
\label{tab:page-mapping}

\renewcommand{\arraystretch}{1.5}
\setlength{\tabcolsep}{10pt}

\begin{tabular}{@{}c>{\raggedright}p{4.2cm}>{\centering\arraybackslash}p{1.5cm}>{\raggedright\arraybackslash}p{2.8cm}@{}}
\toprule
\rowcolor{quantumNavy}
\textcolor{white}{\tabMapCase~\textbf{Case}} &
\textcolor{white}{\tabMapComp~\textbf{Component}} &
\textcolor{white}{\tabMapStage~\textbf{Stage}} &
\textcolor{white}{\tabMapFig~\textbf{Figure}} \\
\midrule
\rowcolor{pageMapTintA}
0 & \texttt{NobelPrizePage} &
\tikz[baseline=-2pt]{\node[circle,fill=stageP1,text=white,font=\tiny\bfseries,inner sep=2pt,minimum size=1.4em]{P1};} &
Fig.~\ref{fig:page1} \\
\rowcolor{white}
1 & \texttt{QuantumBlockchainPage} &
\tikz[baseline=-2pt]{\node[circle,fill=stageP2,text=white,font=\tiny\bfseries,inner sep=2pt,minimum size=1.4em]{P2};} &
Fig.~\ref{fig:page2} \\
\rowcolor{pageMapTintA}
2 & \texttt{QuantumReadyPage} &
\tikz[baseline=-2pt]{\node[circle,fill=stageP3,text=white,font=\tiny\bfseries,inner sep=2pt,minimum size=1.4em]{P3};} &
Fig.~\ref{fig:page3} \\
\rowcolor{white}
3 & \texttt{SentimentPage} &
\tikz[baseline=-2pt]{\node[circle,fill=stageP4,text=white,font=\tiny\bfseries,inner sep=2pt,minimum size=1.4em]{P4};} &
Fig.~\ref{fig:page4} \\
\midrule
\rowcolor{pageMapTintB}
4 & \texttt{SimulatorSelectionPage} &
\tikz[baseline=-2pt]{\node[circle,fill=stageP6,text=white,font=\tiny\bfseries,inner sep=2pt,minimum size=1.4em]{P6};} &
Fig.~\ref{fig:page6} \\
\rowcolor{white}
5 & \texttt{QPUSelectionPage} &
\tikz[baseline=-2pt]{\node[circle,fill=stageP6,text=white,font=\tiny\bfseries,inner sep=2pt,minimum size=1.4em]{P6};} &
Fig.~\ref{fig:page7} \\
\rowcolor{pageMapTintB}
6 & \texttt{KeyGenerationPage} &
\tikz[baseline=-2pt]{\node[circle,fill=stageP7,text=white,font=\tiny\bfseries,inner sep=2pt,minimum size=1.4em]{P7};} &
Fig.~\ref{fig:page8} \\
\rowcolor{white}
7 & \texttt{CompletionPage} &
\tikz[baseline=-2pt]{\node[circle,fill=stageP7,text=white,font=\tiny\bfseries,inner sep=2pt,minimum size=1.4em]{P7};} &
\textcolor{gray}{\tabMapDash~---} \\
\bottomrule
\end{tabular}
\end{table*}

\textbf{Pages 1--2 (Context~$\rightarrow$~Understanding).} Participants explore the physical foundations of superconducting quantum computing (Page~1) and the quantum threat model (Page~2). The interface visualizes the Quantum Vulnerability Index: Shor's algorithm compromises RSA, ECDSA, and Diffie--Hellman; Grover's algorithm reduces effective security margins for SHA-256; AES-256 retains adequate security~\cite{Cherbal2024IoTSecurity}. Post-quantum alternatives (lattice-based signatures, KEMs, hash-based constructions) are introduced. This stage addresses \textit{Security, Privacy \& Forensics}.

\textbf{Pages 3--5 (Participation~$\rightarrow$~Reflection).} Participants provide informed consent (Page~3), input qualitative sentiment on quantum computing (Page~4), and vote to prioritize quantum-safe technologies---PQC, QKD, hash-based cryptography, QRNG, ZKPs---across blockchain-relevant dimensions (Page~5). This models technology prioritization under uncertainty and illustrates ecosystem coordination relevant to \textit{Blockchain for Metaverse \& Digital Twins}.

\textbf{Pages 6--7 (Decision~$\rightarrow$~Outcome).} Participants compare quantum execution environments---trapped-ion, superconducting, neutral-atom, and simulator-based---across performance, availability, and sustainability dimensions~\cite{GYONGYOSI201951,AbuGhanem2025SuperconductingReview,Saffman2019NeutralAtoms}, surfacing the tradeoffs along which these architectures differ, drawing on the simulator and QPU options detailed in Tables~\ref{tab:simulators}--\ref{tab:qpus}. They generate a demonstrative post-quantum artifact (toy LWE-based construction, explicitly labeled as not for production use) with execution provenance metadata linking device selection to cryptographic output. This stage addresses \textit{Performance, Scalability \& Sustainability Issues}.

\textbf{SDG alignment.} All stages contribute to SDG~4, SDG~9, and SDG~16 through technical literacy, resilient infrastructure awareness, and trustworthy digital systems. Pages~3--5 emphasize SDG~17 through participatory collaboration; Pages~6--7 address SDG~12 and SDG~13 through infrastructure sustainability informed by quantum computing life-cycle assessment~\cite{Cordier2025EnvironmentalLCAQuantum}.

\section{User Interface Workflow}
\label{sec:ui-condensed}

To make the live demonstration reproducible for readers, we summarize the mapping from implemented React pages to conceptual stages in Table~\ref{tab:page-mapping} above. The interface implements the workflow across eight React component pages (case indices 0--7 in \texttt{page.tsx}): \textit{NobelPrizePage}, \textit{QuantumBlockchainPage}, \textit{QuantumReadyPage}, \textit{SentimentPage}, \textit{SimulatorSelectionPage}, \textit{QPUSelectionPage}, \textit{KeyGenerationPage}, and \textit{CompletionPage}.

The infrastructure tradeoff stage (P6) exposes the device options listed in Tables~\ref{tab:simulators}--\ref{tab:qpus}: four simulator backends (SV1, DM1, TN1, CUDA-Q GPU) and four QPU families (IonQ trapped-ion, IQM and Rigetti superconducting, QuEra neutral-atom analog). Participants receive framing on environmental impact (energy consumption profiles, cryogenic cooling requirements) and relevance to blockchain infrastructure (quantum-derived randomness for key generation and consensus).

The artifact-generation stage (P7) produces a demonstrative post-quantum cryptographic artifact comprising a quantum-derived identifier, public verification key, digital signature, and execution metadata. The signature algorithm is a toy LWE-based construction (\texttt{ToyLWE-Quantum-Seeded-Demo}) with deliberately reduced parameters; it carries an explicit educational-use warning and is not intended for production deployment. Complete page-by-page UI descriptions with embedded screenshots appear in Appendix~\ref{app:ui-workflow}.

\section*{Acknowledgment}

The authors thank the participants and organizers of the tutorials at The Web
Conference 2026 \cite{10.1145/3774905.3793916} and IEEE ICBC 2026
\cite{guo2026blockchaininfrastructureintelligentcyberphysicalsocial}, where
\textit{Quantum Futures Interactive} was presented as an interactive demonstration
and subsequently improved based on participant feedback.

\bibliographystyle{ieeetr}
\bibliography{reference}

@misc{liu2026quantumcircuitvisioncostaware,
  author       = {Dongping Liu and Aoyu Zhang and Luyao Zhang},
  title        = {{Quantum Circuit Vision}: Cost-Aware Evaluation of
                  Visual {AI} Agents for Quantum Code Generation},
  year         = {2026},
  howpublished = {arXiv preprint arXiv:2607.10057},
  url          = {https://arxiv.org/abs/2607.10057}
}

@misc{zhang2026quantumcinemainteractivecinematic,
  author       = {Aoyu Zhang and Dongping Liu and Luyao Zhang},
  title        = {{Quantum Cinema}: An Interactive Cinematic Exploration
                  of Quantum Computing Hardware via Generative World Models},
  year         = {2026},
  howpublished = {arXiv preprint arXiv:2606.17102},
  url          = {https://arxiv.org/abs/2606.17102}
}

@inproceedings{10.1145/3774905.3793916,
author = {Liu, Dongping and Zhang, Aoyu and Zhang, Luyao},
title = {Quantum-Safe, Efficient, and AI-Enhanced Blockchains for the Web: A Cooperative Tutorial on Quantum Computing, Blockchain Applications, and Data Standards},
year = {2026},
isbn = {9798400723087},
publisher = {Association for Computing Machinery},
address = {New York, NY, USA},
url = {https://doi.org/10.1145/3774905.3793916},
doi = {10.1145/3774905.3793916},
booktitle = {Companion Proceedings of the ACM Web Conference 2026},
pages = {35–38},
numpages = {4},
location = {United Arab Emirates},
series = {WWW Companion '26}
}

@misc{guo2026blockchaininfrastructureintelligentcyberphysicalsocial,
      title={Blockchain Infrastructure for Intelligent Cyber--Physical--Social Systems:Post-Quantum Security, Interoperability, and Trustworthy Data Economies in the Era of Embodied AI}, 
      author={Song Guo and Huawei Huang and Dongping Liu and Aoyu Zhang and Luyao Zhang},
      year={2026},
      eprint={2606.06895},
      archivePrefix={arXiv},
      primaryClass={cs.CR},
      url={https://arxiv.org/abs/2606.06895}, 
}

@misc{qsignai2025,
      title={QSignAI: Quantum-Randomness-Seeded Identity Signatures at the Intersection of AI for Science and Science for AI}, 
      author={Dongping Liu and Aoyu Zhang and Luyao Zhang},
      year={2026},
      eprint={2605.27729},
      archivePrefix={arXiv},
      primaryClass={cs.CR},
      url={https://arxiv.org/abs/2605.27729}, 
}

@article{Carlsen2022SDGStatus,
  author  = {Carlsen, Lars and Br{\"u}ggemann, Ralf},
  title   = {The 17 United Nations’ sustainable development goals: a status by 2020},
  journal = {International Journal of Sustainable Development \& World Ecology},
  volume  = {29},
  number  = {3},
  pages   = {219--229},
  year    = {2022},
  doi     = {10.1080/13504509.2021.1948456},
  url     = {https://doi.org/10.1080/13504509.2021.1948456}
}

@article{Saffman2019NeutralAtoms,
  author  = {Saffman, Mark},
  title   = {Quantum computing with neutral atoms},
  journal = {National Science Review},
  volume  = {6},
  number  = {1},
  pages   = {24--25},
  year    = {2019},
  doi     = {10.1093/nsr/nwy088},
  url     = {https://doi.org/10.1093/nsr/nwy088}
}

@article{AbuGhanem2025SuperconductingReview,
  author  = {AbuGhanem, Muhammad},
  title   = {Superconducting quantum computers: who is leading the future?},
  journal = {EPJ Quantum Technology},
  volume  = {12},
  pages   = {102},
  year    = {2025},
  doi     = {10.1140/epjqt/s40507-025-00405-7},
  url     = {https://doi.org/10.1140/epjqt/s40507-025-00405-7}
}

@article{GYONGYOSI201951,
title = {A Survey on quantum computing technology},
journal = {Computer Science Review},
volume = {31},
pages = {51-71},
year = {2019},
issn = {1574-0137},
doi = {https://doi.org/10.1016/j.cosrev.2018.11.002},
url = {https://www.sciencedirect.com/science/article/pii/S1574013718301709},
author = {Laszlo Gyongyosi and Sandor Imre}
}

@ARTICLE{10310184,
  author={Bach, Benjamin and Keck, Mandy and Rajabiyazdi, Fateme and Losev, Tatiana and Meirelles, Isabel and Dykes, Jason and Laramee, Robert S. and AlKadi, Mashael and Stoiber, Christina and Huron, Samuel and Perin, Charles and Morais, Luiz and Aigner, Wolfgang and Kosminsky, Doris and Boucher, Magdalena and Knudsen, Søren and Manataki, Areti and Aerts, Jan and Hinrichs, Uta and Roberts, Jonathan C. and Carpendale, Sheelagh},
  journal={IEEE Transactions on Visualization and Computer Graphics}, 
  title={Challenges and Opportunities in Data Visualization Education: A Call to Action}, 
  year={2024},
  volume={30},
  number={1},
  pages={649-660},
  doi={10.1109/TVCG.2023.3327378}}

@ARTICLE{9815155,
  author={Yang, Qinglin and Zhao, Yetong and Huang, Huawei and Xiong, Zehui and Kang, Jiawen and Zheng, Zibin},
  journal={IEEE Open Journal of the Computer Society}, 
  title={Fusing Blockchain and AI With Metaverse: A Survey}, 
  year={2022},
  volume={3},
  number={},
  pages={122-136},
  doi={10.1109/OJCS.2022.3188249}}

@article{Cherbal2024IoTSecurity,
  author  = {Cherbal, Samir and Zier, Abdelkader and Hebal, Samia and others},
  title   = {Security in internet of things: a review on approaches based on blockchain, machine learning, cryptography, and quantum computing},
  journal = {The Journal of Supercomputing},
  volume  = {80},
  pages   = {3738--3816},
  year    = {2024},
  doi     = {10.1007/s11227-023-05616-2},
  url     = {https://doi.org/10.1007/s11227-023-05616-2}
}

@article{Cordier2025EnvironmentalLCAQuantum,
  author  = {Cordier, Sylvain and Thibault, Karl and Arpin, Marie-Luc and Amor, Ben},
  title   = {Scaling up to problem sizes: an environmental life cycle assessment of quantum computing},
  journal = {Quantum Science and Technology},
  volume  = {10},
  pages   = {025058},
  year    = {2025},
  publisher = {IOP Publishing},
  doi     = {10.1088/2058-9565/adc0ba},
  url     = {https://doi.org/10.1088/2058-9565/adc0ba}
}

@article{ahmad2023engineeringsoftwaresystemsquantum,
author = {Ahmad, Aakash and Altamimi, Ahmed B. and Aqib, Jamal},
title = {A reference architecture for quantum computing as a service},
year = {2024},
issue_date = {Jul 2024},
publisher = {Elsevier Science Inc.},
address = {USA},
volume = {36},
number = {6},
issn = {1319-1578},
url = {https://doi.org/10.1016/j.jksuci.2024.102094},
doi = {10.1016/j.jksuci.2024.102094},
journal = {J. King Saud Univ. Comput. Inf. Sci.},
month = jul,
numpages = {18}
}

@article{Gonzalez2021AmazonBraket,
  author  = {Gonzalez, Carmen},
  title   = {Cloud based QC with Amazon Braket},
  journal = {Digitale Welt},
  volume  = {5},
  pages   = {14--17},
  year    = {2021},
  doi     = {10.1007/s42354-021-0330-z},
  url     = {https://doi.org/10.1007/s42354-021-0330-z}
}

@ARTICLE{9756233,
  author={Li, Yongkang and Lin, Yanying and Wang, Yang and Ye, Kejiang and Xu, Chengzhong},
  journal={IEEE Transactions on Services Computing}, 
  title={Serverless Computing: State-of-the-Art, Challenges and Opportunities}, 
  year={2023},
  volume={16},
  number={2},
  pages={1522-1539},
  doi={10.1109/TSC.2022.3166553}}

@article{10.1145/3555591,
author = {Williams, Spencer and Jones, Ridley and Reinecke, Katharina and Hsieh, Gary},
title = {An HCI Research Agenda for Online Science Communication},
year = {2022},
issue_date = {November 2022},
publisher = {Association for Computing Machinery},
address = {New York, NY, USA},
volume = {6},
number = {CSCW2},
url = {https://doi.org/10.1145/3555591},
doi = {10.1145/3555591},
journal = {Proc. ACM Hum.-Comput. Interact.},
month = nov,
articleno = {490},
numpages = {22}
}

@article{Bernstein2017PostQuantum,
  author  = {Bernstein, Daniel J. and Lange, Tanja},
  title   = {Post-quantum cryptography},
  journal = {Nature},
  volume  = {549},
  pages   = {188--194},
  year    = {2017},
  doi     = {10.1038/nature23461},
  url     = {https://doi.org/10.1038/nature23461}
}

@techreport{Yaga2018BlockchainOverview,
  author      = {Yaga, Dylan and Mell, Peter and Roby, Nik and Scarfone, Karen},
  title       = {Blockchain Technology Overview},
  institution = {National Institute of Standards and Technology (NIST)},
  number      = {NISTIR 8202},
  year        = {2018},
  month       = oct,
  doi         = {10.6028/NIST.IR.8202},
  url         = {https://doi.org/10.6028/NIST.IR.8202}
}

@article{Budish2025TrustAtScale,
  author  = {Budish, Eric},
  title   = {Trust at Scale: The Economic Limits of Cryptocurrencies and Blockchains},
  journal = {The Quarterly Journal of Economics},
  volume  = {140},
  number  = {1},
  pages   = {1--62},
  year    = {2025},
  publisher = {Oxford University Press},
  doi     = {10.1093/qje/qjae033},
  url     = {https://doi.org/10.1093/qje/qjae033}
}

@INPROCEEDINGS{10048976,
  author={Bavdekar, Ritik and Jayant Chopde, Eashan and Agrawal, Ankit and Bhatia, Ashutosh and Tiwari, Kamlesh},
  booktitle={2023 International Conference on Information Networking (ICOIN)}, 
  title={Post Quantum Cryptography: A Review of Techniques, Challenges and Standardizations}, 
  year={2023},
  volume={},
  number={},
  pages={146-151},
  doi={10.1109/ICOIN56518.2023.10048976}}

@article{Joseph2022PostQuantum,
  author  = {Joseph, Daniel and Misoczki, Rafael and Manzano, M. and others},
  title   = {Transitioning organizations to post-quantum cryptography},
  journal = {Nature},
  volume  = {605},
  pages   = {237--243},
  year    = {2022},
  doi     = {10.1038/s41586-022-04623-2},
  url     = {https://doi.org/10.1038/s41586-022-04623-2}
}

@ARTICLE{11059920,
  author={Wang, Yong and Shahril Ismail, Eddie},
  journal={IEEE Access}, 
  title={A Review on the Advances, Applications, and Future Prospects of Post-Quantum Cryptography in Blockchain and IoT}, 
  year={2025},
  volume={13},
  number={},
  pages={112962-112977},
  doi={10.1109/ACCESS.2025.3584473}}

@misc{cryptoeprint:2025/1626,
      author = {Kigen Fukuda and Shin’ichiro Matsuo and Yuji Suga and Tadahiko Ito},
      title = {The Grand Challenge of {PQC} Migration: Analysis of Modern Blockchain and Intertwined Human Egoisms},
      howpublished = {Cryptology {ePrint} Archive, Paper 2025/1626},
      year = {2025},
      url = {https://eprint.iacr.org/2025/1626}
}

@misc{nistpqc,
  author       = {{National Institute of Standards and Technology (NIST)}},
  title        = {Post-Quantum Cryptography Standardization Project},
  howpublished = {\url{https://csrc.nist.gov/projects/post-quantum-cryptography}},
  note         = {Accessed: 2026-02-16}
}
\appendices


\section{User Interface Workflow and Technical Context}
\label{app:ui-workflow}

This appendix provides a technical description of the eight-page interaction
workflow implemented in \textit{Quantum Futures Interactive}, presented as
seven conceptual stages (P1--P7) in the main text.
The live system implements the workflow across eight React component pages
(case indices 0--7 in \texttt{page.tsx}):
\begin{itemize}
  \item \texttt{NobelPrizePage}
  \item \texttt{QuantumBlockchainPage}
  \item \texttt{QuantumReadyPage}
  \item \texttt{SentimentPage}
  \item \texttt{SimulatorSelectionPage}
  \item \texttt{QPUSelectionPage}
  \item \texttt{KeyGenerationPage}
  \item \texttt{CompletionPage}
\end{itemize}
The seven-stage conceptual grouping in Table~I merges the two device-selection
pages (simulator and QPU) into the ``Infrastructure Tradeoffs'' stage (P6)
and absorbs the \texttt{CompletionPage} into the artifact-generation stage (P7).

The interface presents a structured progression through the scientific
foundations of quantum computation, the implications of quantum algorithms for
cryptographic systems, infrastructure-level decision-making, and the generation
of a demonstrative post-quantum cryptographic artifact.

\subsection{Page 1 -- Scientific Context}
\label{app:page1}

\begin{figure}[htbp]
\centering
\includegraphics[width=0.85\columnwidth]{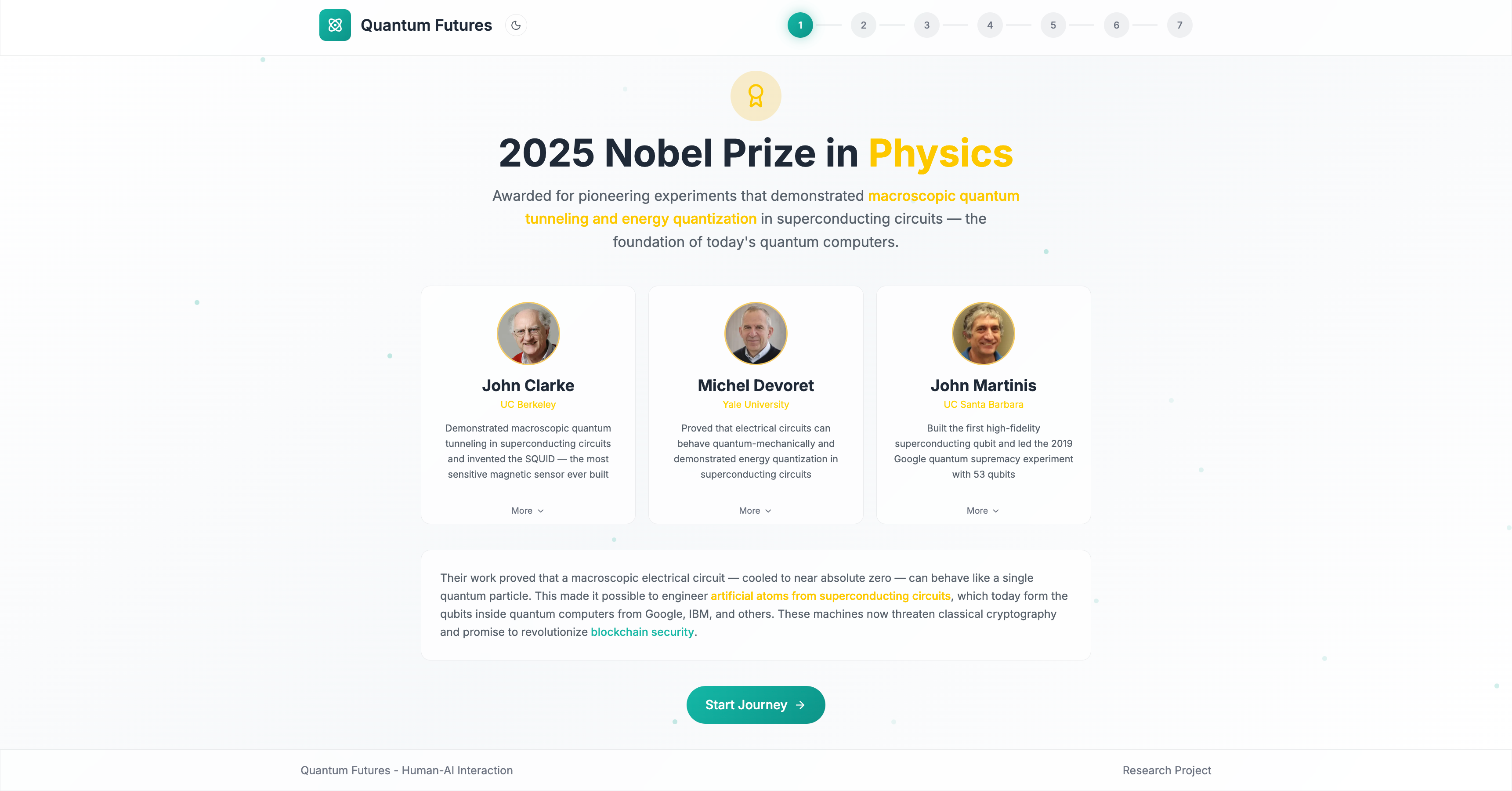}
\caption{Interface Page 1 introducing macroscopic quantum phenomena and
the physical foundations of superconducting quantum computing.}
\label{fig:page1}
\end{figure}

Page 1 (Fig.~\ref{fig:page1}) introduces the physical principles underlying
contemporary quantum computing systems. Experimental demonstrations of
macroscopic quantum tunneling and energy quantization in superconducting
circuits established that engineered electrical systems can exhibit
quantum-mechanical behavior when operated at cryogenic temperatures.
These developments enabled the creation of artificial atoms whose quantized
energy levels can be controlled and measured, forming the basis of
superconducting qubits.

Superconducting qubits are implemented using Josephson junction circuits that
provide nonlinear inductance, allowing isolation of two energy levels to form a
quantum two-level system. Microwave control pulses implement quantum gate
operations through unitary evolution, enabling programmable quantum computation.
The presentation establishes the relationship between advances in quantum
hardware and computational models capable of affecting classical cryptographic
assumptions that underpin distributed trust systems.

\subsection{Page 2 -- Quantum Threat Model and Post-Quantum Cryptographic Transition}
\label{app:page2}

\begin{figure}[htbp]
\centering
\includegraphics[width=0.85\columnwidth]{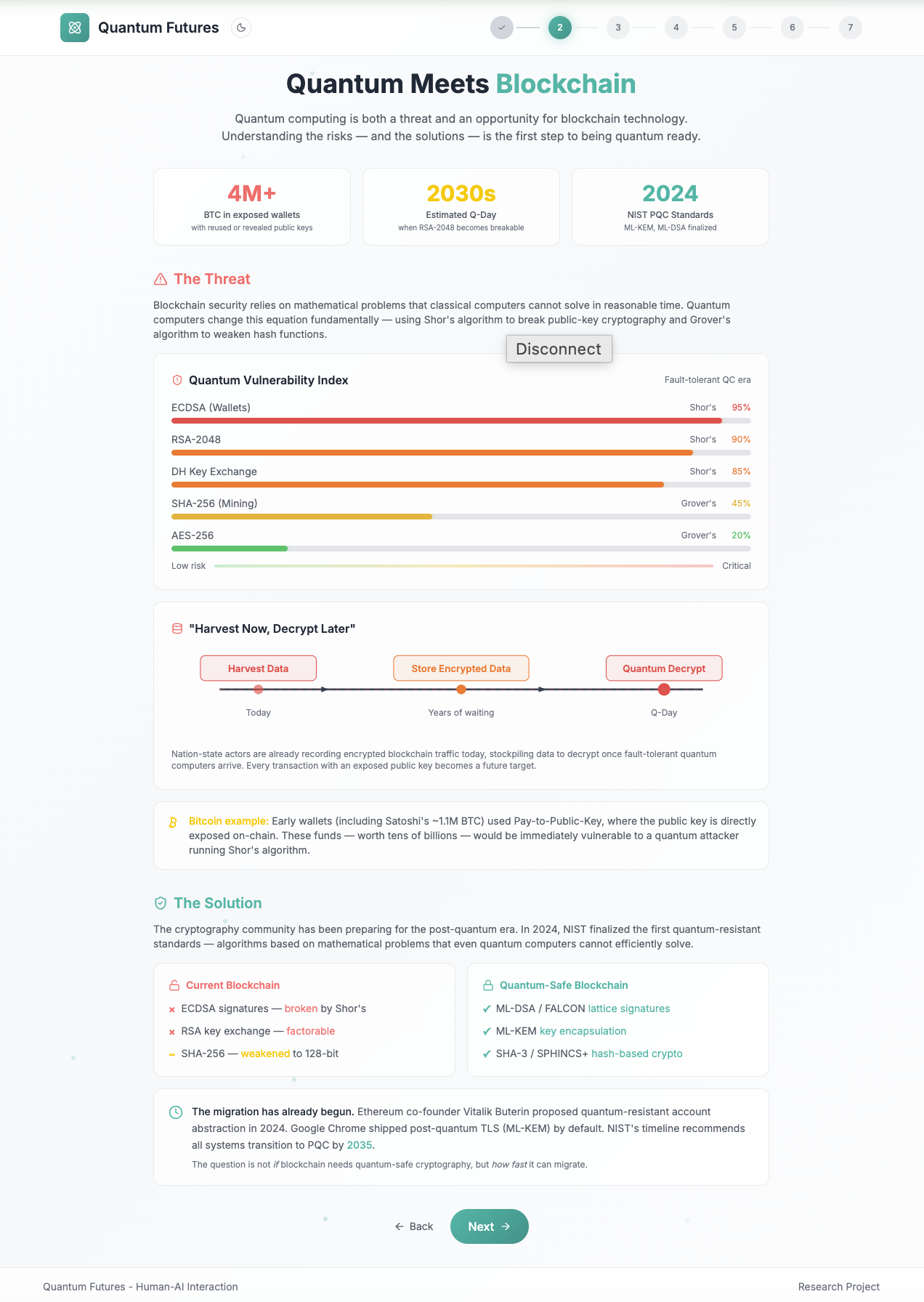}
\caption{Interface Page 2 illustrating the quantum threat model and relative
vulnerability of cryptographic primitives under fault-tolerant quantum computation.}
\label{fig:page2}
\end{figure}

Page 2 (Fig.~\ref{fig:page2}) introduces the impact of quantum algorithms on
widely deployed cryptographic primitives through a visual representation
referred to as the Quantum Vulnerability Index. The interface distinguishes
between two categories of quantum algorithmic effects.

Shor's algorithm enables efficient solutions to integer factorization and
discrete logarithm problems, compromising cryptographic systems such as RSA,
ECDSA, and Diffie--Hellman. Grover's algorithm provides a quadratic speedup
for unstructured search, reducing the effective security margin of symmetric
encryption and hash functions.

The vulnerability indicators correspond to the following technical
interpretations:
\begin{itemize}
\item \textbf{ECDSA:} Vulnerable once public keys are exposed, as quantum
algorithms enable recovery of the corresponding private key.
\item \textbf{RSA:} Authentication loses its security guarantees due to
efficient integer factorization achievable under fault-tolerant quantum
computation.
\item \textbf{Diffie--Hellman:} Key exchange loses forward secrecy in the
presence of quantum adversaries capable of solving discrete logarithm problems.
\item \textbf{SHA-256:} Experiences a reduction in effective security strength
due to quadratic speedups from quantum search algorithms.
\item \textbf{AES-256:} Retains adequate security margins, as quantum attacks
reduce effective key strength but do not compromise the underlying construction.
\end{itemize}

The interface introduces a transition period preceding large-scale
fault-tolerant quantum computation and presents post-quantum cryptographic
alternatives including lattice-based signatures, lattice-based key encapsulation
mechanisms, and hash-based signature constructions.

\subsection{Page 3 -- Participation and Governance Layer}
\label{app:page3}

\begin{figure}[htbp]
\centering
\includegraphics[width=0.85\columnwidth]{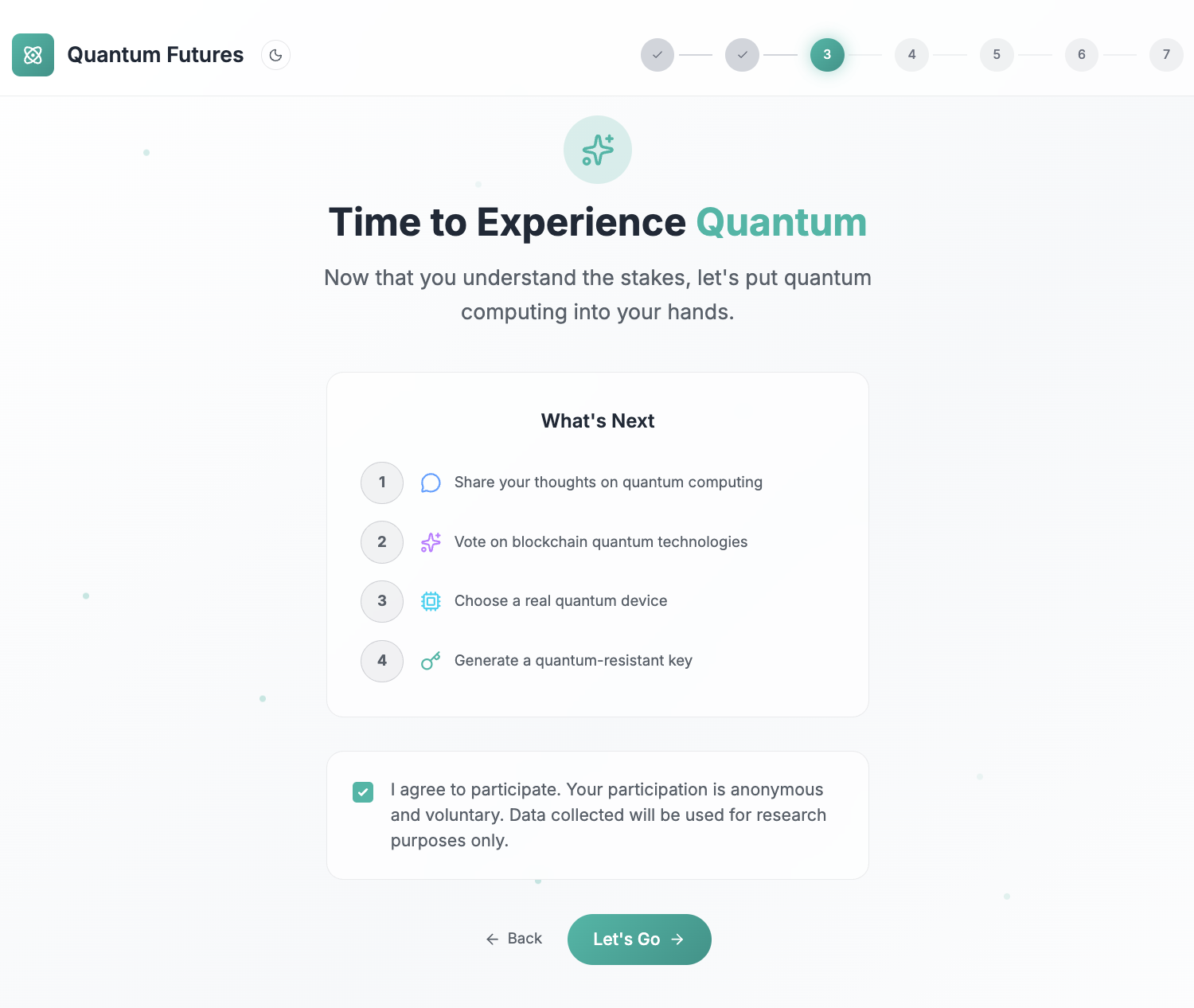}
\caption{Interface Page 3 presenting participation consent and interaction
overview.}
\label{fig:page3}
\end{figure}

Page 3 (Fig.~\ref{fig:page3}) introduces participation and consent mechanisms.
Participation is voluntary and anonymous, and responses are collected for
research purposes. This stage reflects governance considerations relevant to
decentralized infrastructures in which participation contributes to shared
system state while maintaining privacy and transparency.

\subsection{Page 4 -- Public Sentiment Input}
\label{app:page4}

\begin{figure}[htbp]
\centering
\includegraphics[width=0.85\columnwidth]{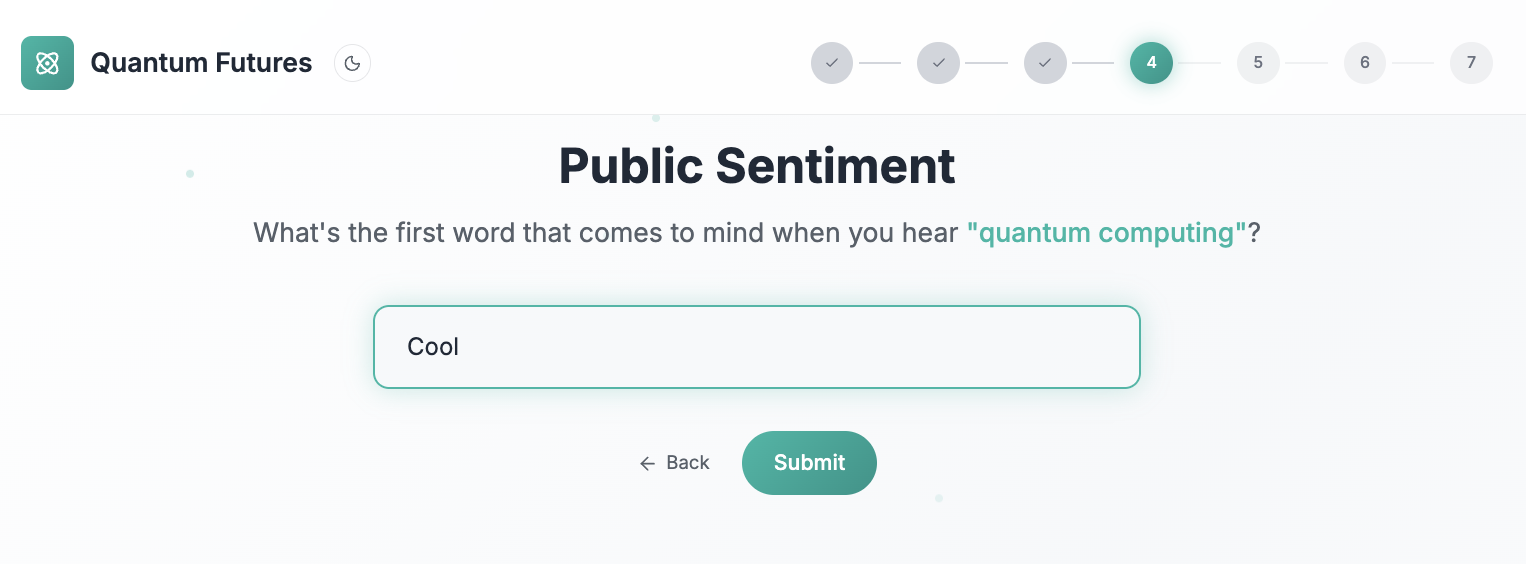}
\caption{Interface Page 4 collecting qualitative public perception of quantum
computing.}
\label{fig:page4}
\end{figure}

Page 4 (Fig.~\ref{fig:page4}) collects a single-word association with quantum
computing. The interaction captures qualitative perception data that is
subsequently aggregated and visualized. Although simple in input structure,
this stage reflects the role of stakeholder perception in technology adoption
and infrastructure transitions.

\subsection{Page 5 -- Sentiment Aggregation and Technology Prioritization}
\label{app:page5}

\begin{figure}[htbp]
\centering
\includegraphics[width=0.85\columnwidth]{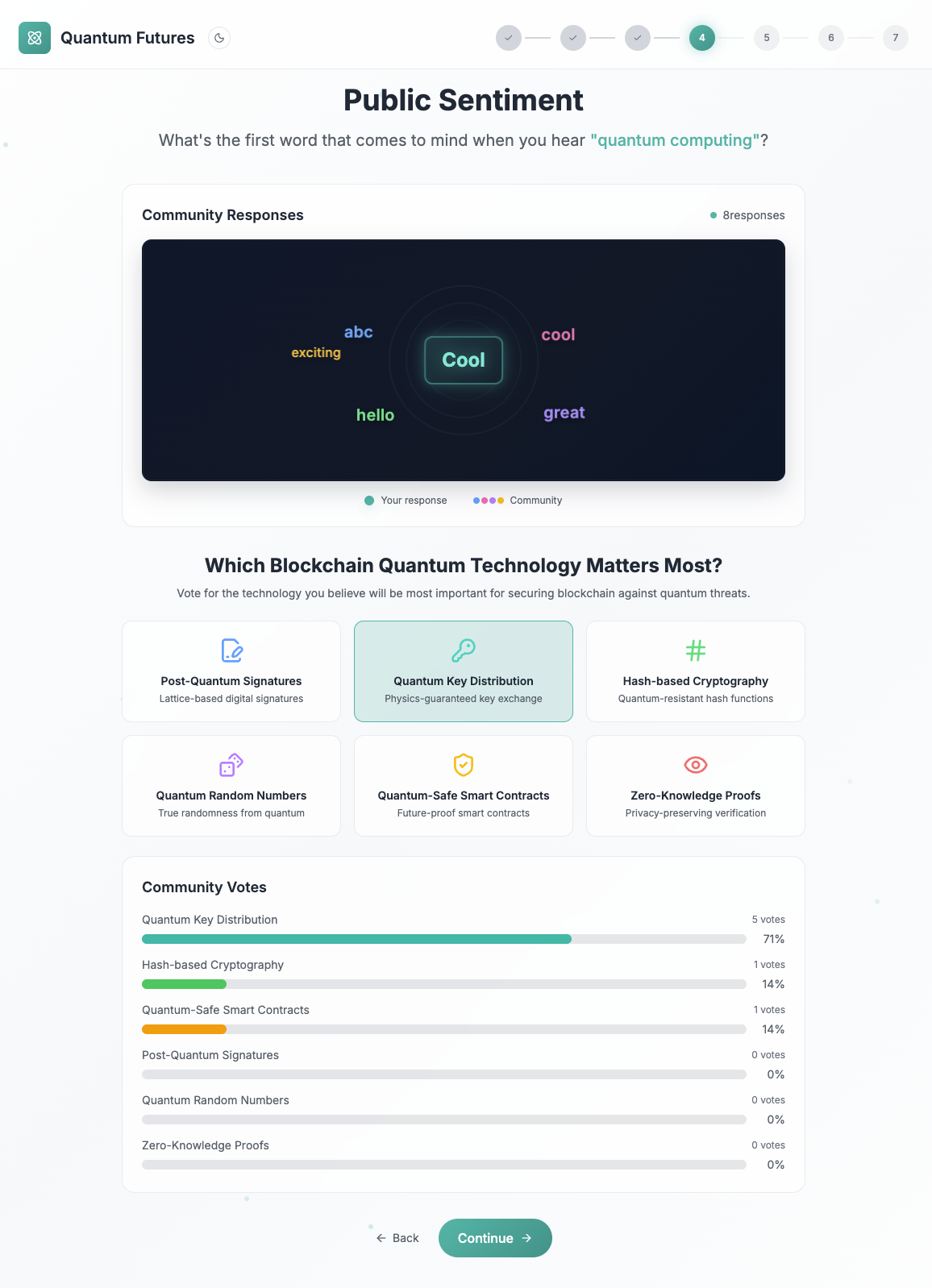}
\caption{Interface Page 5 visualizing aggregated sentiment and enabling voting
on quantum-relevant technologies.}
\label{fig:page5}
\end{figure}

Page 5 (Fig.~\ref{fig:page5}) presents the aggregation of participant
responses collected in the previous stage and introduces a structured voting
mechanism over quantum-related technologies relevant to blockchain and
distributed system security.

The visualization displays collective sentiment as a spatial aggregation of
qualitative inputs, allowing participants to observe how individual perceptions
contribute to a shared representation of community understanding.

The voting process models technology prioritization under conditions of
uncertainty, reflecting how distributed communities evaluate tradeoffs between
security guarantees, deployability, infrastructure cost, and interoperability
with existing systems. In blockchain ecosystems, protocol upgrades frequently
require broad consensus; therefore, understanding collective perception and
prioritization provides insight into potential adoption pathways for
post-quantum technologies. Participants select one of: Post-Quantum Signatures,
Quantum Key Distribution (QKD), Hash-based Cryptography, Quantum Random Number
Generation (QRNG), Quantum-Safe Smart Contracts, or Zero-Knowledge Proofs (ZKPs).

\subsection{Page 6 -- Quantum Simulator Selection and CUDA-Q Integration}
\label{app:page6}

\begin{figure}[htbp]
\centering
\includegraphics[width=0.85\columnwidth]{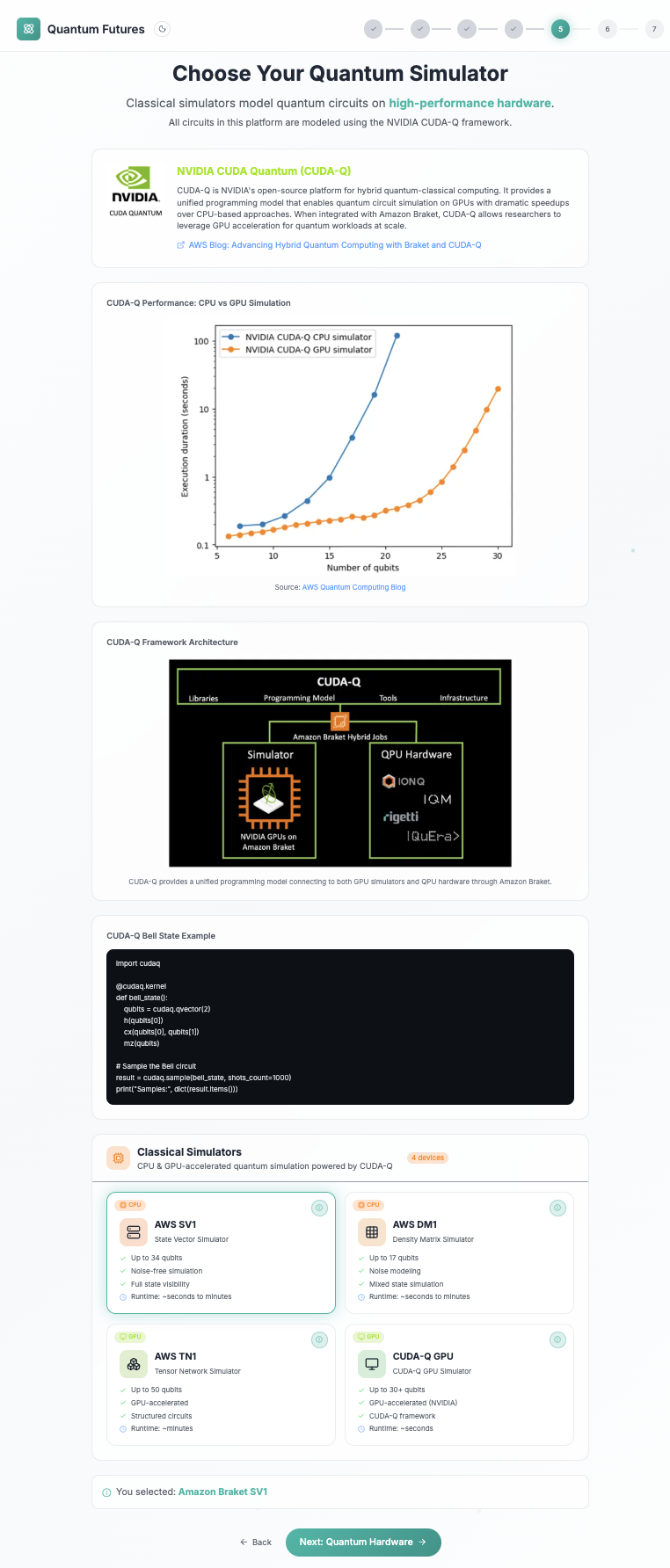}
\caption{Interface Page 6 presenting quantum simulator selection with NVIDIA
CUDA-Q framework integration and GPU-accelerated simulation options.}
\label{fig:page6}
\end{figure}

Page 6 (Fig.~\ref{fig:page6}) introduces infrastructure-level decision-making
through the selection of classical quantum simulation environments. The page
explains the NVIDIA CUDA Quantum (CUDA-Q) framework---an open-source platform
for hybrid quantum-classical computing---and presents available simulator
options.

\subsubsection{CUDA-Q Framework Integration}

All quantum circuits in the platform are modeled using the NVIDIA CUDA Quantum
framework. The page explains how CUDA-Q provides kernel-based quantum
programming (\texttt{@cudaq.kernel}), GPU-accelerated simulation backends
(\texttt{nvidia}, \texttt{nvidia-mgpu}, \texttt{tensornet}), performance
advantages of GPU vs.~CPU simulation, and integration with Amazon Braket
Hybrid Jobs.

\begin{figure}[htbp]
\centering
\includegraphics[width=0.85\columnwidth]{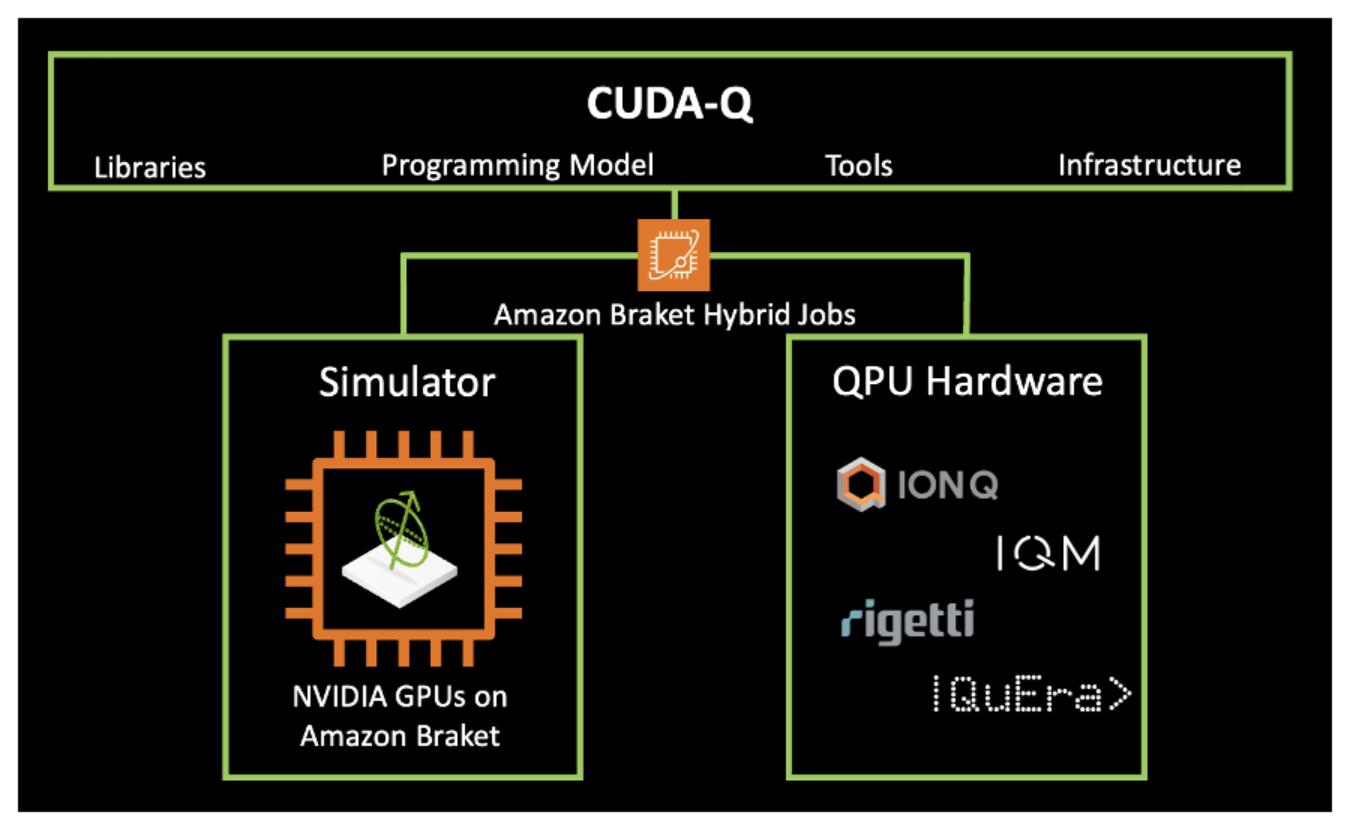}
\caption{CUDA-Q framework architecture showing kernel-based quantum
programming, GPU-accelerated backends, and Amazon Braket Hybrid Jobs integration.}
\label{fig:cuda-q-framework}
\end{figure}

\begin{figure}[htbp]
\centering
\includegraphics[width=0.85\columnwidth]{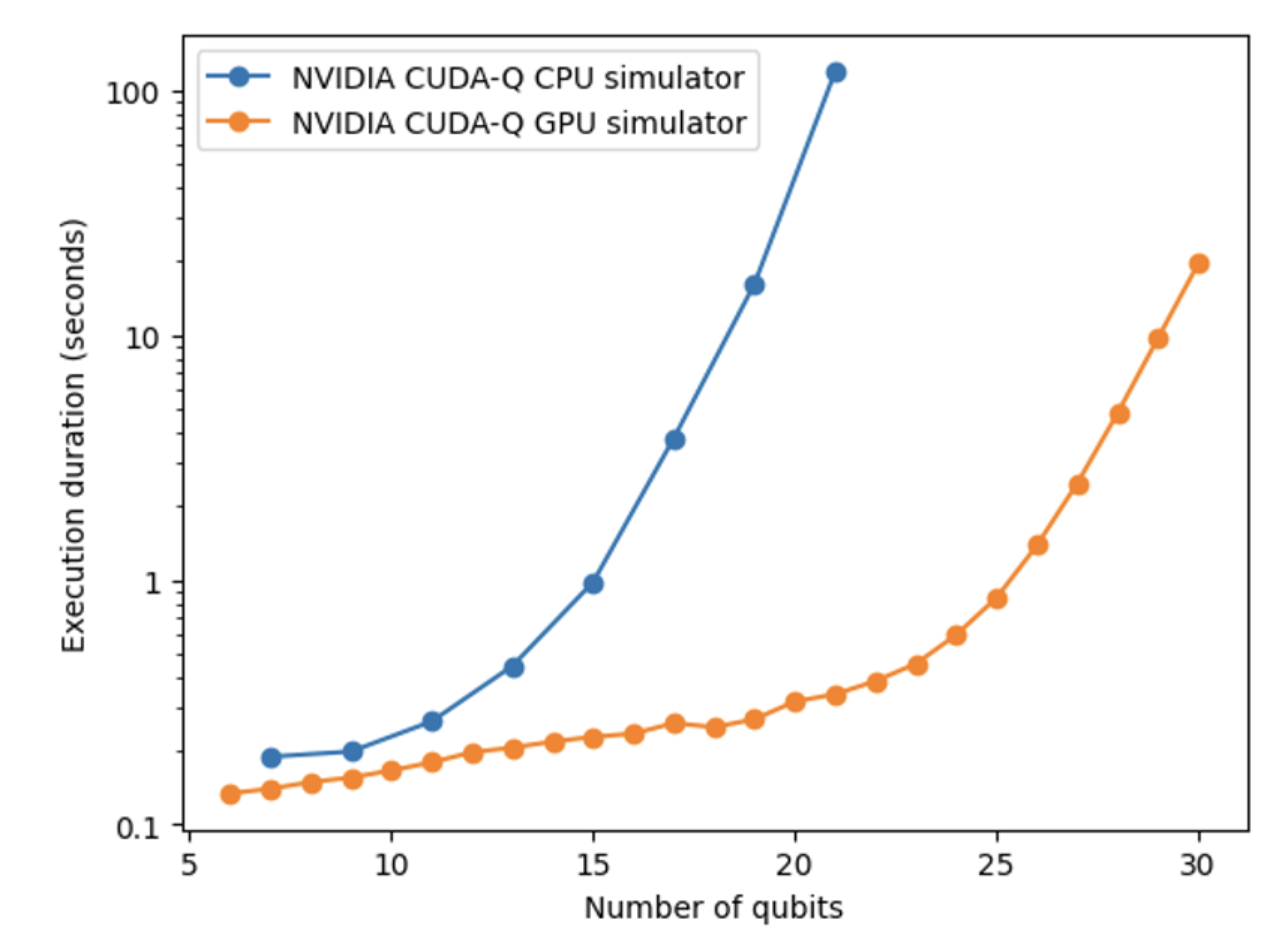}
\caption{CUDA-Q GPU simulator performance comparison: CPU vs.~GPU simulation
execution times for varying qubit counts.}
\label{fig:cuda-q-gpu}
\end{figure}

\subsubsection{Available Simulators}

The simulator alternatives shown to participants are summarized in Table~\ref{tab:simulators} in the main paper. Classical simulators provide deterministic execution environments suitable for verification and debugging, but they do not generate physical randomness because outcomes are computed rather than measured.

\subsection{Page 7 -- Quantum Processing Unit (QPU) Selection}
\label{app:page7}

\begin{figure}[htbp]
\centering
\includegraphics[width=0.85\columnwidth]{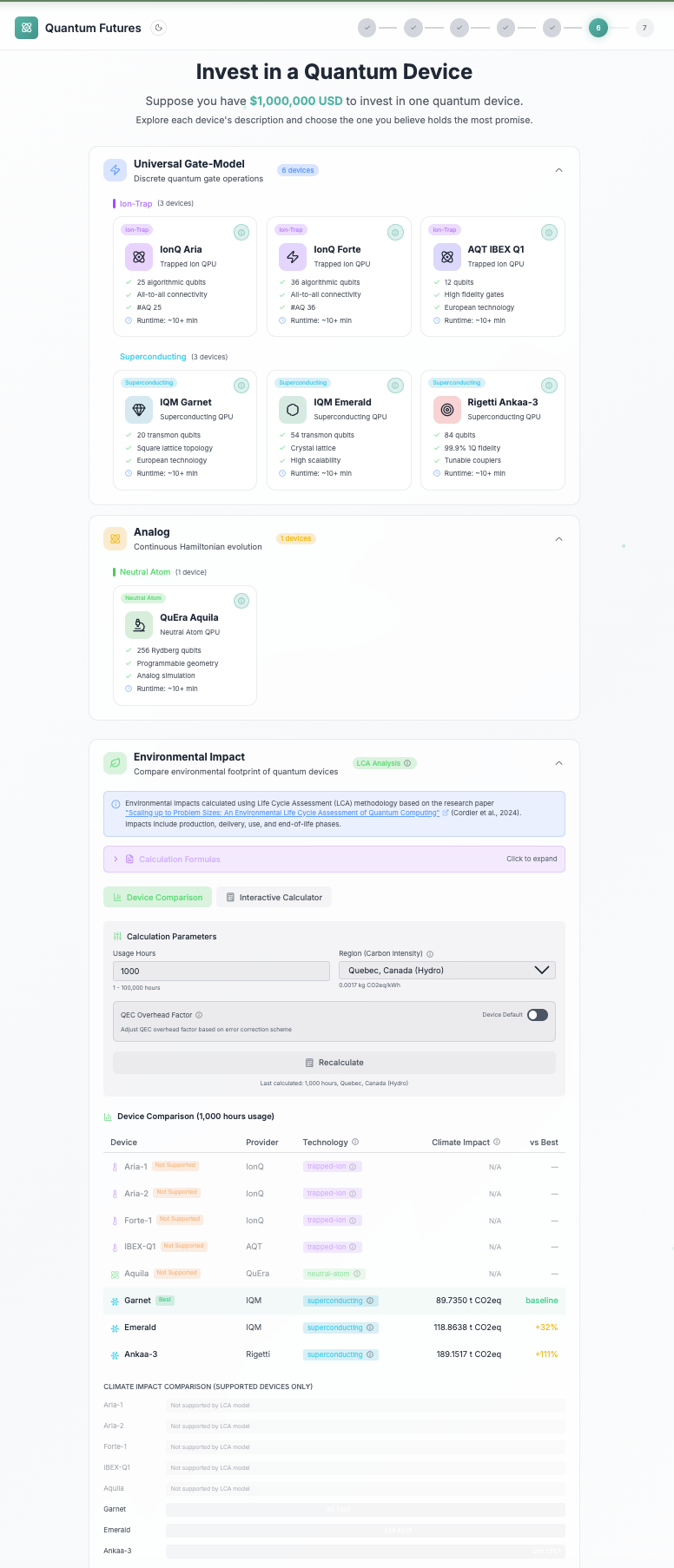}
\caption{Interface Page 7 presenting Quantum Processing Unit selection with
investment framing across trapped-ion, superconducting, and neutral-atom
architectures.}
\label{fig:page7}
\end{figure}

Page 7 (Fig.~\ref{fig:page7}) introduces real quantum hardware decision-making through an investment scenario framing. The interface exposes the diversity of contemporary quantum computing implementations by allowing participants to compare computational models, physical realizations, and operational characteristics.

Because quantum computation depends directly on hardware realization, device selection affects execution fidelity, measurement statistics, scalability, and operational constraints that propagate into higher-level computational and cryptographic workflows.

\subsubsection{Quantum Device Alternatives}

The hardware architecture options are summarized in Table~\ref{tab:qpus} in the main paper.

The interface distinguishes between three execution models: classical
simulation, gate-based quantum computation, and analog quantum evolution.
Gate-based quantum processors operate on physical qubits and produce
probabilistic measurement outcomes derived from quantum superposition, enabling
entropy generation and sampling-based computation. Analog quantum systems
simulate physical Hamiltonians directly, enabling scalable representations of
many-body systems while introducing different reproducibility considerations
due to calibration and continuous evolution dynamics.

These architectural differences result in distinct tradeoffs between execution
fidelity, scalability, latency, and operational complexity, which must be
considered when integrating quantum computation into persistent computational
infrastructures.

\subsubsection{Environmental Impact}

In addition to computational characteristics, the interface introduces
environmental impact estimation through life-cycle assessment parameters.
Quantum computing platforms differ substantially in energy consumption profiles.
Superconducting systems require continuous cryogenic cooling at millikelvin
temperatures, resulting in significant operational overhead. Trapped-ion systems
rely on precision laser control and vacuum environments, while neutral-atom
platforms require optical trapping and calibration infrastructure. Classical
simulators, although energy intensive for large simulations, operate within
conventional data center environments and therefore exhibit different
sustainability tradeoffs.

\subsubsection{Relevance to Blockchain Infrastructure}

Quantum device selection becomes directly relevant to blockchain-integrated
infrastructures when quantum computation contributes to cryptographic processes
or identity generation. Quantum-derived randomness used for key generation,
consensus mechanisms, or identity construction depends on physical measurement
outcomes whose statistical properties vary across hardware implementations.
As a result, device calibration, noise characteristics, and measurement fidelity
become part of the trust boundary when quantum-generated outputs are
incorporated into distributed ledgers.

\subsection{Page 8 -- Post-Quantum Artifact Generation and Execution Provenance}
\label{app:page8}

\begin{figure}[htbp]
\centering
\includegraphics[width=0.85\columnwidth]{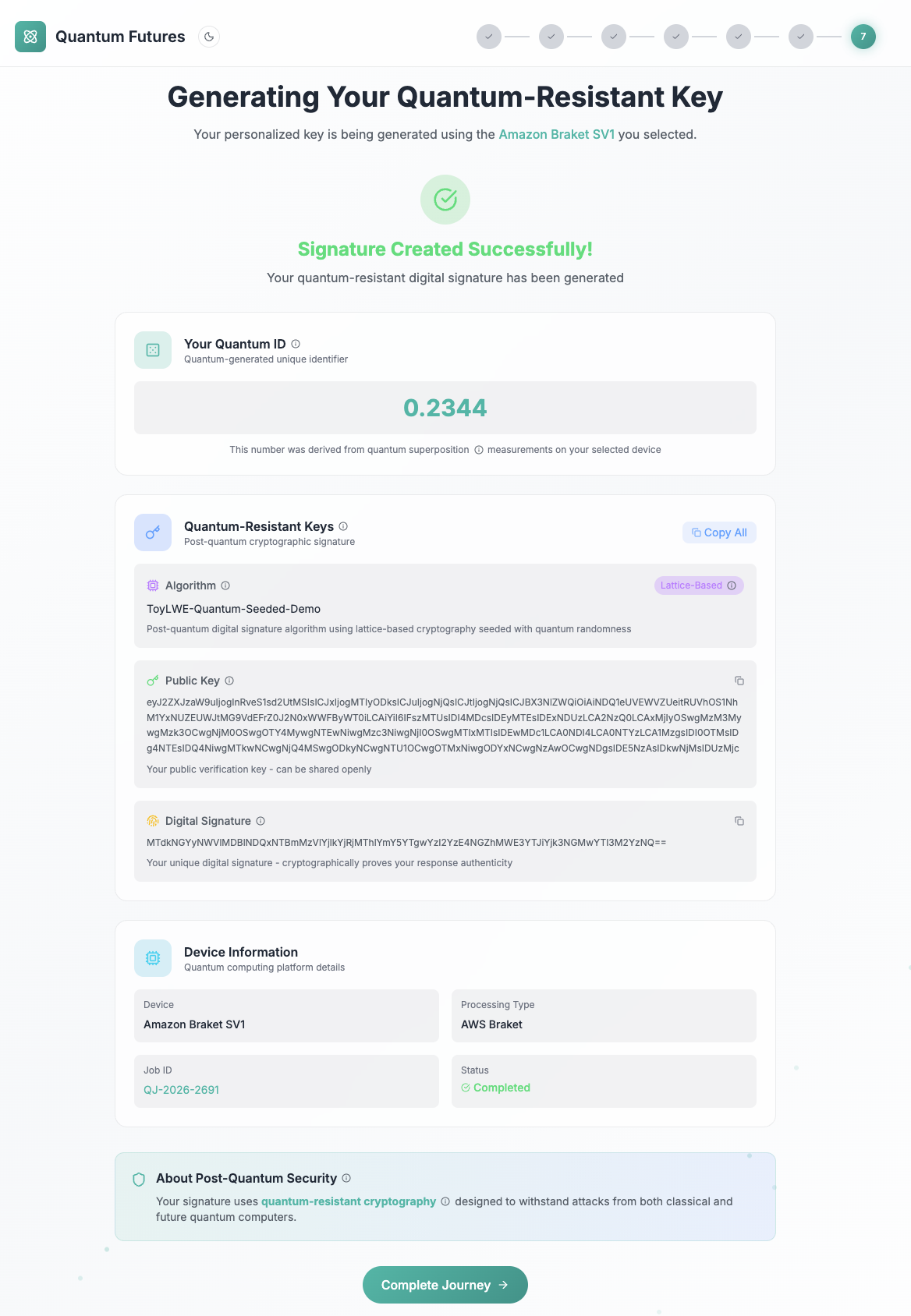}
\caption{Interface Page 8 presenting generated demonstrative post-quantum
cryptographic artifact and execution metadata.}
\label{fig:page8}
\end{figure}

Page 8 (Fig.~\ref{fig:page8}) shows an example output page generated from one
set of selections made on Page 6 or Page 7. The interface presents the
resulting demonstrative post-quantum cryptographic artifact produced after
execution using the selected device and configuration.

The page displays a quantum-derived identifier, public verification key, digital
signature, and execution metadata describing the computational platform, selected
device, and execution status. Quantum measurements are used as a source of
entropy for the cryptographic process, while the signature algorithm represents
a toy LWE-based construction (\texttt{ToyLWE-Quantum-Seeded-Demo}) for
demonstration purposes only, not intended to provide security guarantees
equivalent to standards-grade post-quantum schemes such as ML-DSA or Falcon.
The implementation uses deliberately reduced parameters for educational
demonstration and is explicitly labeled as not for production use.

The inclusion of device information and execution identifiers provides execution
provenance and supports verification of the generated artifact.

\subsection{Implementation Mapping}
\label{app:implementation-mapping}

The mapping between implemented React pages, the seven conceptual stages used in the main text, and the appendix figure references is summarized in Table~\ref{tab:page-mapping} in the main paper.

The progress stepper in the interface displays eight positions (corresponding to
cases 0--7), grouped into seven conceptual stages as presented in Table~I of
the main text.
\end{document}